# Trade Liberalization and Product Innovation: The Dynamic Role of Exporting

Sizhong Sun

College of Business Law and Governance
James Cook University
Queensland, Australia

Email: sizhong.sun@jcu.edu.au

**[Abstract]**

How does trade liberalization affect firms' incentives to innovate? We answer this question using China's WTO accession and a dynamic model of firms' joint export and product-innovation decisions. We find that trade liberalization reduced iceberg trade costs by approximately 13.5 percent in the air-conditioner manufacturing industry, which substantially increased the probability of export and product innovation. Moreover, the response is driven primarily by dynamic rather than static incentives. Two dynamic mechanisms are central: exporting and innovating lead firms to predict more favorable transitions to better productivity states and reduce future entry costs, thereby raising the returns to subsequent innovation. Counterfactual decompositions show that these mechanisms account for the majority of the innovation response. For firms in an intermediate productivity state, the direct static effect (without entry-cost saving and state transition) accounts for only 9.2 percent of the total effect of WTO accession. Shutting down state transition leads to 62.4 percent of the total effect, while shutting down entry-cost saving results in 28.5 percent. The results demonstrate that the innovation effects of trade liberalization are substantially amplified by firms' endogenous dynamic responses, highlighting the importance of accounting for state transitions and forward-looking decisions when evaluating the gains from trade.

* The author thanks Reshad Ahsan, Andreas Moxnes, Jasmine Jiang, Lingfei Lu, Phillip McCalman, Laura Puzzello and participants of the Online Australasian Seminar in International Economics, the 17th Australasian Trade Workshop, the 33rd Annual Conference of Chinese Economics Society Australia, the 2024 Econometric Society Australasia Meeting, the 2025 Econometric Society Winter School in Dynamic Structural Econometrics, and the 2026 Chinese Economists Society China Annual Conference for their comments and suggestions. The author also benefits from the 2022 Econometric Society Summer School in Dynamic Structural Econometrics. All errors remain the author's.

## 1. Introduction

How does trade liberalization affect firms' incentives to innovate? Trade liberalization, namely a reduction in trade barriers, is conventionally understood as a change in firms' access to foreign markets: it lowers the cost of selling abroad and thereby encourages exporting. Moreover, expansion into foreign markets may also change the return to developing new products, thus affecting product innovation.[1] If a larger market helps firms to achieve a higher payoff from an innovation, then trade liberalization incentivizes them to innovate, even when trade policy does not directly target firms' innovation activities. Existing studies frequently confirm such a market-size effect, for example Lileeva & Trefler (2010) and Coelli, Moxnes & Ulltveit-Moe (2022). In addition, the consequences may extend beyond this contemporaneous market-size effect. Exporting and innovating can alter firms' future productivity and the costs of undertaking these activities again, hence creating incentives that operate over time.

This paper studies this interaction between trade liberalization, exporting, and product innovation. We show that a reduction in trade costs, due to trade liberalization, can generate an economically important increase in the probability of product innovation through exporting, and that a substantial part of this effect arises from firms' forward-looking decisions. The key insight is that while exporting raises the contemporaneous marginal return to product innovation, thus encouraging innovation, additionally export and innovation decisions also affect firms' perception of future productivity evolution and future entry costs. Consequently, the effect of trade liberalization on innovation is not fully captured by its contemporaneous effect. Trade liberalization can create a dynamic innovation multiplier, that is, a reduction in trade costs first increases the incentive to export, which then raises the incentive to innovate today and changes the value of innovation and exporting tomorrow.

To separate these different mechanisms, we develop and estimate a dynamic model in which firms jointly choose whether to export and whether to conduct product innovation. Firms are heterogenous, differing in production and innovation capabilities, and make forward-looking decisions about entry, exit, whether to export, and whether to innovate. Product innovation changes the attractiveness of a firm's products and consequently its demand, while exporting gives the firm access to a foreign market, subject to iceberg trade costs. Importantly, firms' export and innovation decisions affect the evolution of their future productivity states, which, as predicted by firms, may exhibit first-order stochastic dominance. That is, firms predict that with export and product innovation they have a better chance of moving to a better state in the future. In addition, both exporting and innovation involve entry costs that need not be paid again once firms have undertaken the activity. These features generate three distinct channels through which trade liberalization can affect product innovation.

The first channel is a static contemporaneous complementarity between exporting and innovation. For an exporting firm, an improvement in product attractiveness can be sold not only in the domestic market, but also in the foreign market. The marginal benefit of product innovation is therefore greater for an exporter than for an otherwise identical non-exporter. Consequently, a reduction in trade costs that induces firms to export immediately incentivizes them to conduct product innovation. This channel is closely related to the market-size effect, namely access to larger markets can raise firms' incentives to innovate, but here the relevant market expansion is endogenous: trade liberalization changes firms' export decisions, which in turn changes the return to innovation. This contemporaneous effect corroborates with the findings of Bustos (2011), Lileeva & Trefler (2010), Atkeson & Burstein (2010), and Aw, Roberts & Xu (2011).

[1] Definition of product innovation can be found at OECD (2005).

The second is a dynamic state-transition channel. Firms understand that their current decisions contain information about future state. In the model (and data), exporting and innovation are associated with a higher predicted probability of moving to a better productivity state in the next period. The prediction of improved future productivity consequently affects the current value of exporting and innovation. This mechanism differs from the conventional interpretation of learning by exporting.[2] The relevant object is not necessarily an ex post causal improvement in productivity generated by exporting; rather, firms make an ex ante prediction of the transition probabilities when making current decisions. The prediction of a more favorable future state can itself increase the current incentive to export or innovate.

The third is a dynamic option-value channel. Both exporting and innovation require substantial entry costs. Once a firm has exported previously, however, it can avoid paying the export entry cost in subsequent periods. Therefore, exporting today creates an option value by lowering the cost of continuing the activity in the next period. This consideration affects firms' current choice of exporting, even when the immediate profit from exporting is relatively small. Similarly, the same logic that generates a dynamic incentive to export also applies to product innovation. Hence, trade liberalization can stimulate innovation not only because exporting improves the marginal benefit of product innovation today, but also because entering export markets changes the future value of firms' choices. The dynamic benefit of option value (saving of entry costs) corroborates with the finding of Roberts & Tybout (1997) and Das et al. (2007) regarding export, and we additionally extend to innovation.

We quantify these mechanisms using firm-level data on Chinese air-conditioner manufacturing firms from 2000 to 2007. China's accession to the World Trade Organization (WTO) in December 2001 provides a large trade-liberalization shock that reduced barriers faced by Chinese firms in serving foreign markets.[3] The setting is particularly useful for studying the interaction between exporting and product innovation because the air-conditioner manufacturing industry features sufficient exporting and product innovation activities, and the data contain information on domestic sales, export sales, new-product value, intermediate inputs, employment, and fixed assets, allowing the export and innovation decisions to be jointly studied. Besides, China's WTO accession was a national policy initiative, rather than a policy targeted at the air-conditioner industry, providing a natural source of variation in the trade environment faced by the firms in the sample.

---

[2] Among others, see Bernard & Jensen (1999; 2004), Blalock & Gertler (2004), De Loecker (2007, 2013). Nevertheless, not all studies find evidence of learning by exporting, for example Clerides et al. (1998).

[3] A number of existing studies have explored the impacts of China's WTO accession from different dimensions, for example Bloom et al. (2015), Lu and Yu (2015), Feng et al. (2017), Lin and Long (2020), Liu and Ma (2020), and Dai et al. (2020, 2021).

Our structural estimation finds that China's WTO accession reduced firms' iceberg trade costs by around 13.5 percent. This reduction in trade costs substantially promoted exporting activity and, consequently, incentivized firms to undertake product innovation. In particular, the estimation of transition probability matrices largely confirms the first-order stochastic dominance. That is, firms ex ante predicted to have a better chance of moving to a higher state if they innovated and exported than if not, and have a better chance of moving upward if they were already at a higher state. For the dynamic option value, our estimation points at substantially large export and innovation entry costs, confirming the importance of avoiding entry costs in driving the innovation and export activities.

We conduct counterfactual analysis, where we shut down the dynamic transition and option value, to isolate the three channels and quantify their relative contribution to the overall effect of trade liberalization on the probability of product innovation and export. While the magnitude depends on the starting state, our counterfactual analysis finds a modest static effect, and in contrast substantially large dynamic effects of state transition and option value. For example, starting from a medium-high state, the static effect alone (shutting down both dynamic channels) contributes only around 9 percent to the overall effect. In contrast, shutting down state transition (namely entry cost saving + static effect) contributes around 63 percent, while shutting down the entry cost saving (namely state transition + static effect) contributes around 28 percent. Thus, the innovation response to trade liberalization is not simply a static response to a larger market. It reflects firms' optimal intertemporal adjustment to a change in the profitability of exporting.

The empirical analysis also highlights an important challenge in estimating these mechanisms. Firms in the data are not a random sampling from the underlying population. Potential entrants decide whether to enter the market, while incumbent firms decide whether to remain or exit. These decisions are themselves functions of firms' unobserved capabilities and expected future profits. Consequently, the observed firms are subject to both endogenous sample selection and attrition. These problems are particularly important in a dynamic setting because firms' decisions to export and innovate depend on the evolution of their productivity states. We therefore explicitly incorporate entry and exit into the structural model and use the model-implied selection process in estimation. This allows the export and innovation decisions to be estimated, while accounting for the fact that the set of firms observed in the data is itself an equilibrium outcome.

The paper contributes to existing literature from three aspects. First, it identifies a mechanism through which trade liberalization affects innovation that operates through firms' export decisions. Existing work has established important links between international trade and innovation, including mechanisms based on market size, competition, comparative advantage, and knowledge spillovers.[4] Other work has explored the relationship between exporting and firms' innovation decisions.[5] This study instead focuses on the joint dynamic decision problem: trade liberalization changes export profitability; exporting changes the return to product innovation; and both decisions alter the value of future choices. The contribution is therefore not to claim that trade liberalization always raises innovation, but to identify the economic conditions under which the export response to trade liberalization generates an innovation response.

Second, the paper shows that the interaction between export and innovation has important dynamic implications. Given a trade liberalization shock, the contemporaneous complementarity between export and product innovation is only one component of the response. Firms also take into account the predicted transition of their productivity state and the future savings in entry costs generated by undertaking these activities today. These mechanisms connect the literature on trade and innovation with the literature on dynamic investment and sunk-cost decisions.[6] In particular, the option-value (saving of entry costs) mechanism extends the dynamic interpretation of sunk export-entry costs (Das et al., 2007; Roberts & Tybout, 1997) to product innovation.

Third, the paper develops a structural framework for estimating these mechanisms while accounting for endogenous entry and exit. The framework combines static pricing and level of production innovation decisions (intensive margins) with forward-looking decisions concerning whether to export and whether to innovate (extensive margins). This distinction is important because the object of interest is not simply the reduced-form correlation between exporting and innovation. The aim is to recover the economic primitives that determine firms' optimal responses to a change in trade costs and then use those primitives to decompose the innovation response into its distinct channels.

Our analysis has implications for how the gains from trade liberalization should be interpreted. Trade policy can affect firms through margins that are not directly targeted by the policy. A reduction in trade barriers can increase exports, but the resulting expansion in market opportunities can also change firms' incentives to develop new products. Moreover, when export and innovation decisions are forward-looking, the long-run response can exceed the immediate response implied by a static model. The innovation effects of trade liberalization therefore depend not only on the size of the foreign market or the elasticity of trade with respect to trade costs, but also on the extent to which exporting and innovation are complementary, how firms' productivity states evolve, and how large the sunk costs of these activities are.

---

[4] See, for example, Melitz and Redding (2021) for a discussion on how international trade (export) can promote innovation.

[5] See, for example, among others, Salomon & Shaver (2005), Ederington and McCalman (2008), Atkeson and Burstein (2010), Damijan et al. (2010), Lileeva & Trefler (2010), Lin and Lin (2010), Bustos (2011), Bratti and Felice (2012), Egger and Keuschnigg (2015), Olabisi (2017), Aghion, Bergeaud, Lequien, & Melitz (2018), Blyde et al. (2018), Impullitti and Licandro (2018), Ayllón & Radicic (2019), Dai et al. (2020), De Fuentes, Niosi & Peerally (2021) and Coelli, Moxnes and Ulltveit-Moe (2022).

[6] See, for example, Dixit & Pindyck (1994).

These mechanisms also clarify the scope of the paper's conclusions. The results do not imply that every reduction in trade barriers must increase innovation. The innovation response depends on the economic environment governing the complementarity between exporting and innovation, the productivity transition process, and the persistence and magnitude of entry costs. In environments where exporting does not substantially increase the return to innovation, where exporting does not improve firms' prediction of state transition, or where innovation has little persistence in its associated costs, the dynamic innovation multiplier can be correspondingly smaller. The contribution of the paper is therefore a mechanism-level explanation of when and why trade liberalization can generate innovation, rather than a claim that the effect is universal.

The remainder of the paper is organized into seven sections. Section two presents the analytical framework, where we develop an economic model that features both static and dynamic decision making, and describes the estimation strategy. In section three, we discuss the data, variables, and reduced-form evidence. Section four reports and discusses the estimation results. Section five conducts counterfactual analysis, including decomposing the effects of trade liberalization through the three different channels. In section six, we make various robustness checks, and section seven discusses external validity. Section eight concludes.

## 2. Analytical Framework

### *2.1 Economic model*

In each period, firms engage in a multi-stage decision making process, as described in Figure 1. Upon entering the market, randomly-sampled firms draw production ($\lambda_1$) and innovation ($\lambda_2$) capability endowments from an exogenous distribution with cumulative distribution function (CDF) of $G(\lambda_1, \lambda_2)$. At stage one, entry firms decide whether to enter the market by comparing the value of entry (the present value of all expected future profits) with an outside option. If they decide not to enter the market, they are not observed in the data.

At stage two, firms face shocks, either from the demand side (e.g., a decrease of consumers' disposable income) or from the supply side (e.g., an increase of input price), and decide whether to exit. If they decide to exit, they collect a liquidation value, normalized to 0, and are not observed in the data. Should they decide to stay, they then make decisions on whether to innovate and/or export at stage three. At stages four and five, firms choose an optimal level of product innovation and set prices for both domestic and export markets.[7]

Firms have perfect information in the current period, including the shocks at stage two. However, their information on the states in future periods is imperfect, and as such they form ex ante beliefs on the transition of the states (namely they predict the law of transition) when making dynamic decisions. As will be discussed in more details below, decisions at stages four and five are static, while decisions at stages one, two and three are dynamic.[8]

**The *equilibrium*** is a set of optimal decisions on entry, exit, whether to innovate and/or export, optimal degree of product innovation, and prices in domestic and export markets, together with cut-off capabilities, that clear the markets and firms' beliefs are correct.[9]

Figure 1. The Decision Tree

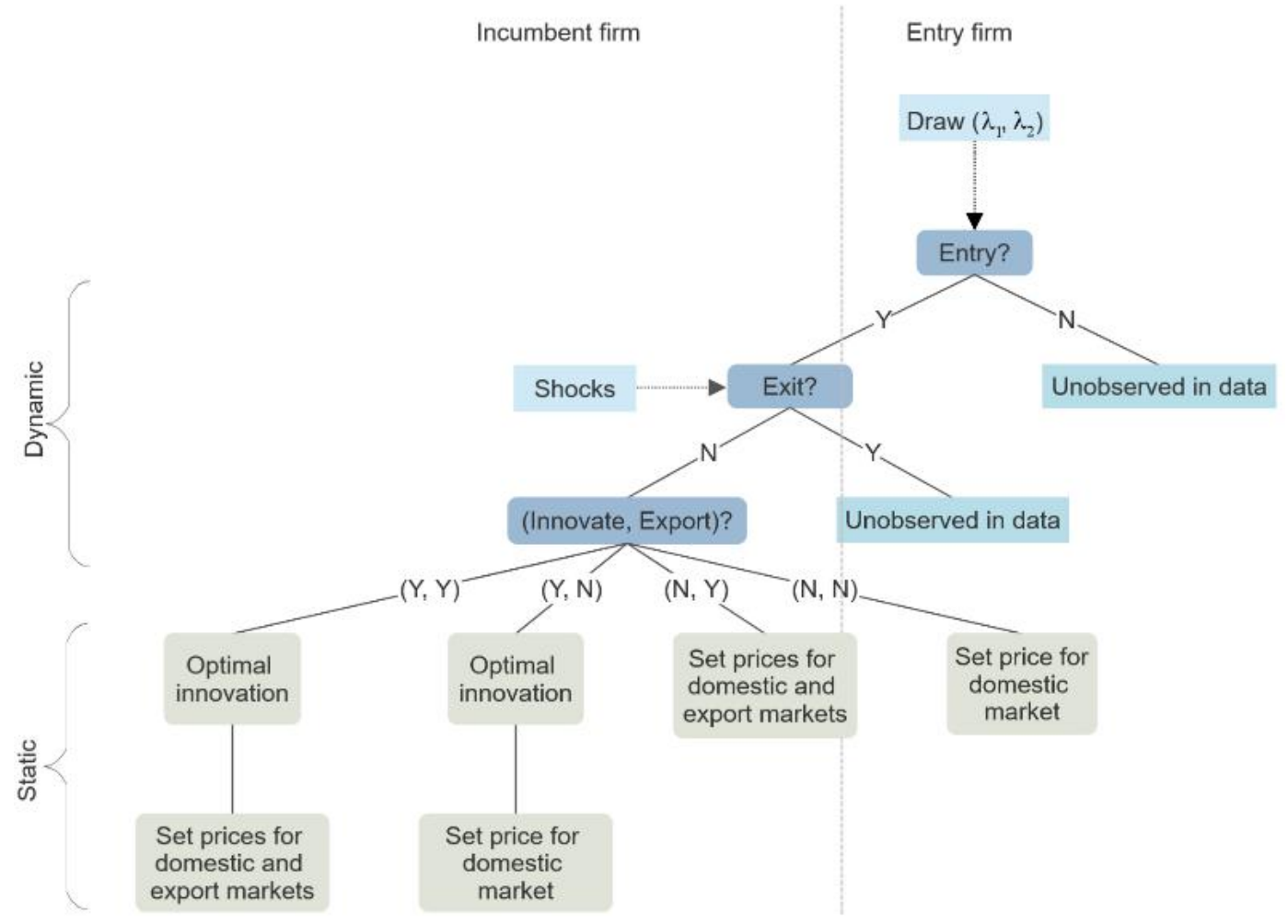

The demand in the domestic market is $q = Ap^{1/\rho - 1}$, where $\rho$ is the constant-elasticity-of-substitution (CES) preference parameter ($0 < \rho < 1$); $q$ is the quantity of goods that is a CES aggregation of new and existing products, namely $q \equiv [q_o^\rho + (\zeta q_n)^\rho]^{1/\rho}$ where the subscripts $o$ and $n$ denote existing and new products respectively,[10] we set the CES aggregation parameter to be $\rho$ by choosing the numeraire appropriately, and $\zeta$ measures the attractiveness of new products, relative to existing products; $A$ is aggregate demand, namely a representative consumer's income divided by a price index; and $p$ represents price.[11] Similarly, in the foreign market, the demand is $\tilde{q} = \tilde{A}\tilde{p}^{1/\tilde{\rho}-1}$, where tilde denotes the foreign market and $\tilde{q} \equiv [\tilde{q}_o^{\rho} + (\zeta\tilde{q}_n)^{\rho}]^{1/\rho}$.

[7] Note our modelling features both ex ante and ex post firm heterogeneity.

[8] That is, decisions on extensive margins are dynamic and decisions on intensive margins are static. It is implicitly assumed that what matters for the future is whether firms do it (extensive margins), rather than how much firms do it (intensive margins).

[9] Being correct means firms' predicted law of transition is consistent with data. Besides, in the long run, the value of entry equals the entry cost.

[10] Note despite we do not frame our modelling in terms of multi-product firms, we are actually modelling multi-product firms with two types of products, the existing and new products. In addition,

Production incurs a fixed cost ($f$), due to facilities used in production. Using these facilities, firms can then combine one unit of labor with $m$ units of intermediate inputs to produce $s$ units of existing products. The term $s$ measures labor productivity, and $s = \lambda_1\psi$ where $\lambda_1$ and $\psi$ are the unobserved and observed components of labor productivity respectively. Alternatively, to produce new products, firms pay an additional fixed cost of innovation ($f_n$), and combine one unit of labor with $m$ units of intermediate inputs to produce $\lambda_2 s$ units of existing products. Let $\eta \equiv \frac{q_n}{q_o}$, which measures the level of product innovation. Then, the marginal cost of production can be written as $c = \frac{(\lambda_2+\eta)(w+m)}{\lambda_2(1+\zeta^{\rho}\eta^{\rho})^{1/\rho}s}$.

Firms' per-period profit from the domestic market is then $\pi = (p - c)q - f - [f_n + f_n^e(1 - \chi_{1,-1})]\chi_1$, where $\pi$ denotes profit; $\chi_1$ takes a value of one if the firm conducts product innovation and zero otherwise; $\chi_{1,-1}$ is the firm's product innovation status in the previous period; $f_n$ and $f_n^e$ are the fixed and entry costs of product innovation, respectively. If the firm has conducted innovation in the previous period, it can avoid the entry costs.[12] To export, firms need to pay a fixed and entry export costs ($f_e$ and $f_e^e$, respectively) and an iceberg trade cost ($\tau$). Export profit can then be written as $\tilde{\pi} = (\tilde{p} - \tau c)\tilde{q} - [f_e + f_e^e(1 - \chi_{2,-1})]$, where $\chi_{2,-1}$ is the firm's export status in the previous period; and similar to innovation, it saves on the entry cost if it has exported in the previous period.[13]

---

$q_o$ and $q_n$ can themselves be a CES aggregation of other multiple sub-types of existing and new products, respectively.

[11] Such a demand function can be obtained by maximizing a CES utility function, subject to a budget constraint.

[12] More generally, as long as a firm innovates in the past, it can save the innovation entry cost, namely its innovation entry cost in year *t* is $f_n^e\left(1 - \mathbb{1}(\sum_{j=0}^{t}\chi_{1j} > 0)\right)$ where $\mathbb{1}(\cdot)$ denotes an indicator function taking a value of 1 if its argument is true and $\chi_{1j}$ takes a value of 1 if the firm innovates in year *j* and 0 otherwise. Note this assumption gives rise to a dynamic benefit of innovation (and similarly for export), namely if a firm innovates this period, it can save entry cost in the next period.

[13] Similar to the innovation entry cost, one can specify a more general export entry cost in year *t* as $f_e^e\left(1 - \mathbb{1}(\sum_{j=0}^{t}\chi_{2j} > 0)\right)$ where $\chi_{2j}$ equals 1 if a firm exports in year *j* and 0 otherwise.

At stage five, firms set optimal prices as a markup over marginal cost of production, namely $p = \frac{c}{\rho}$ and $\tilde{p} = \frac{\tau c}{\tilde{\rho}}$ in the domestic and export markets, respectively. Accordingly, their per-period optimal profits are $\pi = (1 - \rho)\rho^{\frac{\rho}{1-\rho}} A c^{\frac{\rho}{\rho-1}} - f - [f_n + f_n^e(1 - \chi_{1,-1})]\chi_1$ and $\tilde{\pi} = (1 - \tilde{\rho})\rho^{\frac{\tilde{\rho}}{1-\tilde{\rho}}} \tilde{A} \tau^{\frac{\tilde{\rho}}{\tilde{\rho}-1}} c^{\frac{\tilde{\rho}}{\tilde{\rho}-1}} - [f_e - f_e^e(1 - \chi_{2,-1})]$. Hence, at stage four, firms' optimal level of product innovation, should they decide to innovate, is $\eta = \lambda_2^{\frac{1}{1-\rho}} \zeta^{\frac{\rho}{1-\rho}}$.

Therefore, firms' optimal per-period profits from domestic and export markets at stage four are $\pi(\chi_1 = 0) = (1 - \rho)\rho^{\frac{\rho}{1-\rho}} A(w + m)^{\frac{\rho}{\rho-1}} \psi^{\frac{\rho}{1-\rho}} \lambda_1^{\frac{\rho}{1-\rho}} - f$, $\pi(\chi_1 = 1) = (1 - \rho)\rho^{\frac{\rho}{1-\rho}} A(w + m)^{\frac{\rho}{\rho-1}} \psi^{\frac{\rho}{1-\rho}} \left(1 + \lambda_2^{\frac{\rho}{1-\rho}} \zeta^{\frac{\rho}{1-\rho}}\right) \lambda_1^{\frac{\rho}{1-\rho}} - f - [f_n + f_n^e(1 - \chi_{1,-1})]$,

$\tilde{\pi}(\chi_1 = 0) = (1 - \tilde{\rho})\tilde{\rho}^{\frac{\tilde{\rho}}{1-\tilde{\rho}}} \tau^{\frac{\tilde{\rho}}{\tilde{\rho}-1}} \tilde{A}(w + m)^{\frac{\tilde{\rho}}{\tilde{\rho}-1}} \psi^{\frac{\tilde{\rho}}{1-\tilde{\rho}}} \lambda_1^{\frac{\tilde{\rho}}{1-\tilde{\rho}}} - [f_e + f_e^e(1 - \chi_{2,-1})]$, and $\tilde{\pi}(\chi_1 = 1) = (1 - \tilde{\rho})\rho^{\frac{\tilde{\rho}}{1-\tilde{\rho}}} \tau^{\frac{\tilde{\rho}}{\tilde{\rho}-1}} \tilde{A}(w + m)^{\frac{\tilde{\rho}}{\tilde{\rho}-1}} \psi^{\frac{\tilde{\rho}}{1-\tilde{\rho}}} \left(1 + \lambda_2^{\frac{\rho}{1-\rho}} \zeta^{\frac{\rho}{1-\rho}}\right)^{\frac{\tilde{\rho}}{1-\tilde{\rho}} \frac{1-\rho}{\rho}} \lambda_1^{\frac{\tilde{\rho}}{1-\tilde{\rho}}} - [f_e + f_e^e(1 - \chi_{2,-1})]$. Examining the per-period profit functions, we can obtain the static benefit of a reduction of iceberg trade cost, summarized in Proposition 1 and proved in the appendix.

**Proposition 1**: Static benefit of a reduction of iceberg trade cost
A decrease of the iceberg trade cost immediately incentivizes firms to innovate, should they export.

Let $\chi_2$ denote whether a firm exports, taking a value of one if it does and zero otherwise. At stage three, if a firm decides to export (innovate), it needs to pay entry costs, which are waived if it has exported (innovated) in the previous period, due to re-utilizing facilities from the past period. In addition, decisions at stage three affect the evolution of state variable ($\frac{\psi}{w+m}$). Therefore, the stage three decisions are dynamic. In making decisions, firms form beliefs about the evolution of state variables. That is, they predict the law of transition.

**Firms' *beliefs***. For the future shocks, which affect $A$ and $\tilde{A}$, firms understand that there is a probability of $1 - \sigma$ $(0 < \sigma < 1)$ that adversary shocks will occur, resulting in exit in the next period. Conditional on staying in the market, firms forecast their next period's $A$ and $\tilde{A}$ by this period's $A$ and $\tilde{A}$.[14] For the state variable ($\frac{\psi}{w+m}$), firms understand that it follows a finite-state Markov process, conditional on entry and stay in the market, where the transition probability depends on today's innovation and export decisions. Firms predict the transition probability matrices based on their past experience and other available information. In equilibrium, firms' prediction of the law of transition is correct.

[14] Akin to an adaptive expectation.

Given firms' beliefs, we can obtain their value functions. The value of choosing $(\chi_1, \chi_2) = (0,0)$, normalized by the fixed cost of production, is $v_{(0,0)} \equiv \frac{1}{f}\check{v}_{(0,0)} = (1-\rho)\rho^{\frac{\rho}{1-\rho}}\frac{1}{f}A(w+m)^{\frac{\rho}{\rho-1}}\psi^{\frac{\rho}{1-\rho}}\lambda_1^{\frac{\rho}{1-\rho}} - 1 + \delta\sigma\mathbb{E}v_{(0,0)}$, where $\check{v}$ and $v$ denote the original and normalized value functions, respectively; $\delta$ is the discount rate and $\mathbb{E}v_{(0,0)}$ is the expected continuation value where the expectation is taken over the evolution of $\left(\frac{\psi}{w+m}\right)$, conditional on the choice of $(0,0)$ and stay in the market. Similarly, the choice value of $(\chi_1, \chi_2) = (0,1)$ is $v_{(0,1)} \equiv \frac{1}{f}\check{v}_{(0,1)} = (1-\rho)\rho^{\frac{\rho}{1-\rho}}\frac{1}{f}A(w+m)^{\frac{\rho}{\rho-1}}\psi^{\frac{\rho}{1-\rho}}\lambda_1^{\frac{\rho}{1-\rho}} + (1-\tilde{\rho})\tilde{\rho}^{\frac{\tilde{\rho}}{1-\tilde{\rho}}}\tau^{\frac{\tilde{\rho}}{\tilde{\rho}-1}}\frac{1}{f}\tilde{A}(w+m)^{\frac{\tilde{\rho}}{\tilde{\rho}-1}}\psi^{\frac{\tilde{\rho}}{1-\tilde{\rho}}}\lambda_1^{\frac{\tilde{\rho}}{1-\tilde{\rho}}} - 1 - \left[\frac{f_e}{f} + \frac{f_e^e}{f}(1-\chi_{2,-1})\right] + \delta\sigma\mathbb{E}v_{(0,1)}$. The choice value of $(\chi_1, \chi_2) = (1,0)$ is $v_{(1,0)} \equiv \frac{1}{f}\check{v}_{(1,0)} = (1-\rho)\rho^{\frac{\rho}{1-\rho}}\frac{1}{f}A(w+m)^{\frac{\rho}{\rho-1}}\psi^{\frac{\rho}{1-\rho}}\left(1+\lambda_2^{\frac{\rho}{1-\rho}}\zeta^{\frac{\rho}{1-\rho}}\right)\lambda_1^{\frac{\rho}{1-\rho}} - 1 - \left[\frac{f_n}{f} + \frac{f_n^e}{f}(1-\chi_{1,-1})\right] + \delta\sigma\mathbb{E}v_{(1,0)}$. The choice value of $(\chi_1, \chi_2) = (1,1)$ is $v_{(1,1)} \equiv \frac{1}{f}\check{v}_{(1,1)} = (1-\rho)\rho^{\frac{\rho}{1-\rho}}\frac{1}{f}A(w+m)^{\frac{\rho}{\rho-1}}\psi^{\frac{\rho}{1-\rho}}\left(1+\lambda_2^{\frac{\rho}{1-\rho}}\zeta^{\frac{\rho}{1-\rho}}\right)\lambda_1^{\frac{\rho}{1-\rho}} + (1-\tilde{\rho})\tilde{\rho}^{\frac{\tilde{\rho}}{1-\tilde{\rho}}}\tau^{\frac{\tilde{\rho}}{\tilde{\rho}-1}}\frac{1}{f}\tilde{A}(w+m)^{\frac{\tilde{\rho}}{\tilde{\rho}-1}}\psi^{\frac{\tilde{\rho}}{1-\tilde{\rho}}}\left(1+\lambda_2^{\frac{\rho}{1-\rho}}\zeta^{\frac{\rho}{1-\rho}}\right)^{\frac{\tilde{\rho}}{1-\tilde{\rho}}\frac{1-\rho}{\rho}}\lambda_1^{\frac{\tilde{\rho}}{1-\tilde{\rho}}} - 1 - \left[\frac{f_e}{f} + \frac{f_e^e}{f}(1-\chi_{2,-1})\right] - \left[\frac{f_n}{f} + \frac{f_n^e}{f}(1-\chi_{1,-1})\right] + \delta\sigma\mathbb{E}v_{(1,1)}$. Note the normalization by fixed cost of production is without loss of generality and aids identification in the empirical estimations. Due to this normalization, later the parameter estimates of fixed and entry costs of export and innovation measure those relative to fixed cost of production. Examining the value functions, we can obtain the dynamic benefits of a reduction of iceberg trade cost, summarized in Proposition 2 and proved in the appendix.

**Proposition 2**: Dynamic benefits of a reduction of iceberg trade cost

(1) Saving of entry costs: If firms start to export and innovate after a reduction of iceberg trade cost, saving of the entry costs of export and innovation is a dynamic benefit.

(2) State transition: A reduction of iceberg trade cost brings in an additional dynamic benefit, should firms start to export and innovate and their ex ante belief of transition of productivity states is as follows:

(2.1) Export and innovation help them move to a higher productivity state, namely the transition probability matrix of $\left(\frac{\psi}{w+m}\right)$ when firms choose $(\chi_1, \chi_2) = (1,1)$ first-order stochastically dominates that of $(\chi_1, \chi_2) = (0,0)$;

(2.2) At a higher productivity state, they have a better chance of moving upward, namely the transition probability distribution of $\left(\frac{\psi}{w+m}\right)$ at a higher productivity state first-order stochastically dominates that of a lower productivity state.

The dynamic benefit of entry cost saving is a direct consequence of assumption, while the dynamic benefit of moving to a better productivity state hinges on firms' ex ante belief being monotone. One then naturally wonders how big the entry costs are and whether the transition probability matrices exhibit first-order stochastic dominance in reality. These questions are empirical, which we later will estimate from data.

Given the value functions, firms' optimal choices are as follows: $\chi_1 = \mathbb{1}(v_{(1,0)} > v_{(0,0)})\mathbb{1}(v_{(1,0)} > v_{(0,1)})\mathbb{1}(v_{(1,0)} \geq v_{(1,1)}) + \mathbb{1}(v_{(1,1)} > v_{(0,0)})\mathbb{1}(v_{(1,1)} > v_{(0,1)})\mathbb{1}(v_{(1,1)} > v_{(1,0)})$, and $\chi_2 = \mathbb{1}(v_{(0,1)} > v_{(0,0)})\mathbb{1}(v_{(0,1)} > v_{(1,0)})\mathbb{1}(v_{(0,1)} \geq v_{(1,1)}) + \mathbb{1}(v_{(1,1)} > v_{(0,0)})\mathbb{1}(v_{(1,1)} > v_{(1,0)})\mathbb{1}(v_{(1,1)} > v_{(0,1)})$, where $\mathbb{1}(\cdot)$ is the indicator function, taking a value of one if its argument is true and zero otherwise.

We note two observations here. First, since firms can save on entry costs if they export/innovate previously, they are more likely to export/innovate if they have done so previously. Second, as the per-period profits are monotone increasing functions of firm capability endowments ($\lambda_1$ and $\lambda_2$), we expect $\chi_1$ and $\chi_2$ to be positively correlated. That is, an exporting/innovating firm is more likely to innovate/export respectively. Later we will examine the data regarding these two reduced-form predictions.

At stage two, if the realized adversary shock is substantially large, namely it reduces the per-period profit to a sufficiently large extent such that the value of staying in the market is less than its liquidation value, the firm exits. At stage one, firms' value functions depend on their capability endowments. If the realized capability endowments are such that the value functions are negative, firms do not enter the market. The stage two decision results in an endogenous sample attrition, while the stage one problem leads to an endogenous sample selection.

*2.2 Estimation strategy*

We intend to estimate the impacts of trade liberalization, namely China's WTO accession, on firms' export and product innovation. For this purpose, we implement the following four-step algorithm.

In step one, we first estimate the CES preference parameters, $\rho$ and $\tilde{\rho}$, utilizing the structural relationship between total variable costs ($TVC$) and firms' domestic sales ($R$) and export revenues ($\tilde{R}$) implied by profit maximization at stage five, as follows:

$$TVC = \rho R + \tilde{\rho}\tilde{R} + \varepsilon_1 \quad (1)$$

where we append an error term $\varepsilon_1$ to capture the measurement error. Researchers have previously utilized a similar strategy to identify the CES preference parameters (see for example Aw et al., 2011).[15] Later in the estimations, we first regress total variable costs against domestic sales revenues, restricting the sample to non-exporting firms, to estimate $\rho$. Then using the estimated $\rho$, regress $(TVC - \rho R)$ against export revenues using the sample of exporting firms to estimate $\tilde{\rho}$.

In step two, we then proceed to estimate the impact of trade liberalization (WTO accession) on the iceberg trade cost. Firms' optimal per-period domestic sales and export revenues are $R = \rho^{\frac{\rho}{1-\rho}} A c^{\frac{\rho}{\rho-1}}$ and $\tilde{R} = \tilde{\rho}^{\frac{\tilde{\rho}}{1-\tilde{\rho}}} \tilde{A} c^{\frac{\tilde{\rho}}{\tilde{\rho}-1}} \tau^{\frac{\tilde{\rho}}{\tilde{\rho}-1}}$, where the aggregate demands $A$ and $\tilde{A}$ are consumer income divided by an aggregate price index. In the estimation we will use the constant price gross national income (GNI) per capita divided by the base year price, in China and the world, to measure them.

Hence, $\frac{\tilde{\rho}\tilde{R}^{\frac{\tilde{\rho}-1}{\tilde{\rho}}} Y^{\frac{\rho-1}{\rho}}}{\rho R^{\frac{\rho-1}{\rho}} \tilde{Y}^{\frac{\tilde{\rho}-1}{\tilde{\rho}}}} = \frac{\tilde{P}^{\frac{\tilde{\rho}-1}{\tilde{\rho}}}}{P^{\frac{\rho-1}{\rho}}}\tau$, where $Y$ and $\tilde{Y}$ are the constant price GNI per capita in China and the world, respectively; and $P$ and $\tilde{P}$ are the base year prices in China and the world, respectively. Note with estimates of $\rho$ and $\tilde{\rho}$ from step one and data of $(R, Y, \tilde{R}, \tilde{Y})$, the left-hand-side of the equation is observed. In light of the equation, we can use the following specification to identify the effect of WTO accession:

$$lny = \alpha_0 + \alpha_1 dWTO + \varepsilon_2 \quad (2)$$

[15] In Monte Carlo experiments, Sun and Anwar (2024) find that such a strategy has a high identification power.

where $lny = ln\left(\frac{\tilde{\rho}\tilde{R}^{\frac{\tilde{\rho}-1}{\tilde{\rho}}}Y^{\frac{\rho-1}{\rho}}}{\rho R^{\frac{\rho-1}{\rho}}\tilde{Y}^{\frac{\tilde{\rho}-1}{\tilde{\rho}}}}\right)$; $dWTO$ is a dummy variable, taking a value of one if post-WTO accession and zero otherwise;[16] $\varepsilon_2$ captures measurement errors and other firm level factors; $\alpha_0 = ln\left(\frac{\tilde{P}^{\frac{\tilde{\rho}-1}{\tilde{\rho}}}}{P^{\frac{\rho-1}{\rho}}}\right) + ln\tau_0$ where $\tau_0$ is the pre-WTO iceberg trade cost; and $\alpha_1 = ln\tau_1 - ln\tau_0$ where $\tau_1$ is the post-WTO iceberg trade cost.

The parameter $\alpha_1$ measures change of iceberg trade cost after WTO accession. A similar identification strategy has been used in previous studies. For example, Alessandria and Choi (2014) identify the change of iceberg trade cost from change in the export-to-domestic sales ratio and the relative size of the markets. We expect trade liberalization to reduce the iceberg trade cost, namely a negative estimate of $\alpha_1$. Besides, in equation (2), the calculation of $lny$ uses the estimates of $\rho$ and $\tilde{\rho}$ in step one. To account for the estimation errors brought in from step one, we compute the bootstrap standard errors in estimating equation (2) where we randomly re-sample (with placement) not just the data, but also from the distribution of the estimates of $\rho$ and $\tilde{\rho}$ in step one.

In order to estimate the dynamic decisions on whether to innovate and export, we need to have the probability of exit and transition probability of state variable $\left(\frac{\psi}{w+m}\right)$. Since firms have correct belief (prediction) in equilibrium, we can nonparametrically estimate them as the relative frequency in step three. For the exit probability $(1-\sigma)$, it is the number of exit firms divided by the total number of firms in the sample, where a firm exits if it appears in the current year, but not in the next. That is, $\hat{\sigma} = \frac{1}{N}\sum_{i,t}\left(1 - \mathbb{1}(\Omega_{it})\right)$ where $\Omega_{it}$ is the event that firm $i$ appears in year $t$ but not in year $t+1$, the hat represents estimate, and $N$ denotes the sample size. By the law of large numbers, $\hat{\sigma} \xrightarrow{p} \sigma$ and $\sqrt{N}(\hat{\sigma} - \sigma) \xrightarrow{d} \mathcal{N}[0, \sigma(1-\sigma)]$ as $N \to \infty$.

The state variable $\left(\frac{\psi}{w+m}\right)$ contains $\psi$ and $(w+m)$, where $(w+m)$ equals total wage and intermediate inputs divided by the number of workers. For $\psi$, we first assume it depends on firm size (total number of workers) and capital intensity (fixed assets per worker). We then utilize the principal component analysis (PCA) to extract the principal component from data of firm size and capital intensity. Dividing it by $(w+m)$, we further discretize it into *K* intervals, namely $\left(\frac{\psi}{w+m}\right) \in \{1,2,\cdots,K\}$ where a higher level corresponds to a better state.

---

[16] Note the WTO accession includes a mix of policies, such as tariff cut, that aim at trade liberalization. Existing studies use a same time dummy variable to measure WTO accession (for example Liu & Ma, 2020; Lu & Yu, 2015).

The term $\psi$ is one component of firm productivity, and hence conceptually depends on firm size and capital intensity. Using PCA allows us to reduce the dimensionality. For discretization, later we will set $K = 4$. The measurement of $\psi$ and subsequent state discretization is not innocuous because, as it turns out, the state-transition channel is central to our quantitative conclusions. Hence, we will conduct robustness check on measurement of states from two aspects, namely an alternative discretization ($K = 2$) and an alternative measurement of $\psi$ as total factor productivity (TFP).

Conditional on $(\chi_1, \chi_2)$, the transition probability from one state to another is estimated as the share of firms that move from one state in the current year to another in the next year. That is, conditional on $(\chi_1, \chi_2)$, let $\mu_{j|k}$ denote the probability of transitioning from state $k$ to state $j$, $\mu_{j,k}$ be the probability of being in state $j$ currently and in state $k$ previously, and $\mu_k$ be the probability in state $k$, where $k, j \in \{1,2,\cdots,K\}$. Then $\mu_{j|k} = \frac{\mu_{j,k}}{\mu_k}$. The probability $\mu_{j,k}$ can be estimated as $\hat{\mu}_{j,k} = \frac{1}{\widetilde{N}}\sum_{i,t}\mathbb{1}\left(\left(\frac{\psi}{w+m}\right)_{it} = j, \left(\frac{\psi}{w+m}\right)_{it-1} = k\right)$, where $\widetilde{N}$ denotes the size of sample with $(\chi_1, \chi_2)$ and $\lim\limits_{N\to\infty}\frac{\widetilde{N}}{N} > 0$. Then $\hat{\mu}_{j,k} \xrightarrow{p} \mu_{j,k}$ as $N \to \infty$. Similarly, $\hat{\mu}_k = \frac{1}{\widetilde{N}}\sum_{i,t}\mathbb{1}\left(\left(\frac{\psi}{w+m}\right)_{it-1} = k\right)$, and $\hat{\mu}_k \xrightarrow{p} \mu_k$ as $N \to \infty$. The transition probability is estimated as $\hat{\mu}_{j|k} = \frac{\hat{\mu}_{j,k}}{\hat{\mu}_k}$, where $\hat{\mu}_{j|k} \xrightarrow{p} \mu_{j|k}$ and the standard errors can be computed by the bootstrap method.

Step four estimates firms' decisions on whether to innovate and export, which is a dynamic discrete choice model.[17] For this purpose, we further parameterize $\hat{\lambda}_1 \equiv \frac{1}{f}A\lambda_1^{\frac{\rho}{1-\rho}} \sim e^{\mathcal{N}\left[-\frac{1}{2}ln2, ln2\right]}$, $\hat{\lambda}_2 \equiv \lambda_2^{\frac{\rho}{1-\rho}}\zeta^{\frac{\rho}{1-\rho}} \sim e^{\mathcal{N}\left[-\frac{1}{2}ln2, ln2\right]}$, $\tau^{\frac{\tilde{\rho}}{\tilde{\rho}-1}} f^{\frac{(1-\rho)\tilde{\rho}}{\rho(1-\tilde{\rho})}} \tilde{A} A^{-\frac{(1-\rho)\tilde{\rho}}{\rho(1-\tilde{\rho})}} \equiv e^{\beta_0}$, $\frac{f_e}{f} \equiv e^{\beta_1} + \varepsilon_3$, $\frac{f_n}{f} \equiv e^{\beta_2} + \varepsilon_4$, $\frac{f_e^e}{f} \equiv e^{\beta_3 + \beta_4 IsBig} + \varepsilon_5$, and $\frac{f_e^n}{f} \equiv e^{\beta_5 + \beta_6 IsBig} + \varepsilon_6$, where $\varepsilon_i \sim \mathcal{N}[0,1]$, $i = 3,4,5,6$; $IsBig \in \{0,1\}$ is a dummy variable, taking value 1 if a firm is big; and $\beta \equiv (\beta_0, \beta_1, \beta_2, \beta_3, \beta_4, \beta_5, \beta_6)'$ is the vector of parameters that we will estimate from the data. Note $\hat{\lambda}_1$ and $\hat{\lambda}_2$ are log-normally distributed with mean and variance of 1.

Accordingly, the choice value functions become

$$v_{(0,0)} = (1-\rho)\rho^{\frac{\rho}{1-\rho}}\left(\frac{\psi}{w+m}\right)^{\frac{\rho}{1-\rho}}\hat{\lambda}_1 - 1 + \delta\sigma E v_{(0,0)},$$

[17] Researchers have explored estimations of dynamic discrete choice models extensively, for example, to name a few, Hotz and Miller (1993), Imai et al. (2009), Keane and Sauer (2009), Keane and Wolpin (2009), Aguirregabiria (2010), Aguirregabiria and Mira (2010), Abbring (2010), Arcidiacono and Miller (2011), Bugni and Ura (2019), Abbring and Daljord (2020), and Kalouptsidi et al. (2021).

$$v_{(0,1)} = (1-\rho)\rho^{\frac{\rho}{1-\rho}}\left(\frac{\psi}{w+m}\right)^{\frac{\rho}{1-\rho}}\hat{\lambda}_1 + (1-\tilde{\rho})\tilde{\rho}^{\frac{\tilde{\rho}}{1-\tilde{\rho}}}e^{\beta_0}\left(\frac{\psi}{w+m}\right)^{\frac{\tilde{\rho}}{1-\tilde{\rho}}}\hat{\lambda}_1^{\frac{(1-\rho)\tilde{\rho}}{\rho(1-\tilde{\rho})}} - 1 - \left\{e^{\beta_1} + \varepsilon_3 + \left(e^{\beta_3+\beta_4 IsBig} + \varepsilon_5\right)\left(1-\chi_{2,-1}\right)\right\} + \delta\sigma E v_{(0,1)},$$

$$v_{(1,0)} = (1-\rho)\rho^{\frac{\rho}{1-\rho}}\left(\frac{\psi}{w+m}\right)^{\frac{\rho}{1-\rho}}\hat{\lambda}_1\left(1+\hat{\lambda}_2\right) - 1 - \left\{e^{\beta_2} + \varepsilon_4 + \left(e^{\beta_5+\beta_6 IsBig} + \varepsilon_6\right)\left(1-\chi_{1,-1}\right)\right\} + \delta\sigma E v_{(1,0)}, \text{ and}$$

$$v_{(1,1)} = (1-\rho)\rho^{\frac{\rho}{1-\rho}}\left(\frac{\psi}{w+m}\right)^{\frac{\rho}{1-\rho}}\hat{\lambda}_1\left(1+\hat{\lambda}_2\right) + (1-\tilde{\rho})\tilde{\rho}^{\frac{\tilde{\rho}}{1-\tilde{\rho}}}e^{\beta_0}\left(\frac{\psi}{w+m}\right)^{\frac{\tilde{\rho}}{1-\tilde{\rho}}}\hat{\lambda}_1^{\frac{(1-\rho)\tilde{\rho}}{\rho(1-\tilde{\rho})}}\left(1+\hat{\lambda}_2\right)^{\frac{(1-\rho)\tilde{\rho}}{\rho(1-\tilde{\rho})}} - 1 - \left\{e^{\beta_1} + \varepsilon_3 + \left(e^{\beta_3+\beta_4 IsBig} + \varepsilon_5\right)\left(1-\chi_{2,-1}\right)\right\} - \left\{e^{\beta_2} + \varepsilon_4 + \left(e^{\beta_5+\beta_6 IsBig} + \varepsilon_6\right)\left(1-\chi_{1,-1}\right)\right\} + \delta\sigma E v_{(1,1)}.$$

The known parameters in the value functions, from estimations in the previous steps, are $(\rho, \tilde{\rho}, \sigma, \Sigma)$ where $\Sigma$ represents the set of conditional transition probability matrices of discretized $\left(\frac{\psi}{w+m}\right)$, and we set the discount rate $(1-\delta)$ to 0.1 (and 0.05 and 0.01 in robustness check).

In step four, first, we draw $D = 1000$ $\hat{\lambda}_1$ and $\hat{\lambda}_2$ from the standard log-normal distributions and $D = 1000$ $\varepsilon_3$, $\varepsilon_4$, $\varepsilon_5$ and $\varepsilon_6$ from the standard normal distributions.[18] Conditional on $(\rho, \tilde{\rho}, \sigma, \Sigma, \delta)$, one can compute the choice values for each pair of $\left(\hat{\lambda}_1, \hat{\lambda}_2\right)$ with a guess of $\beta$, namely $\left(v^d_{(0,0)}, v^d_{(0,1)}, v^d_{(1,0)}, v^d_{(1,1)}\right)$, $d = 1, \dots, D$, where we utilize a nested fixed-point algorithm in the computation. Such a nested fixed-point algorithm has been utilized in computing value functions in other similar setups, pioneered by Rust (1987). Then, we can simulate the four choice probabilities, conditional on firm entry at stage one and stay at stage two. For example, $\widehat{Pr}(\chi_1 = 1, \chi_2 = 1 | entry, stay) = \frac{1}{\sum_{d=1}^{D}\mathbb{1}\left(v^d_{(0,0)} \geq 0\right)}\sum_{d=1}^{D}\frac{1}{1+exp\left(v^d_{(0,0)}-v^d_{(1,1)}\right)+exp\left(v^d_{(0,1)}-v^d_{(1,1)}\right)+exp\left(v^d_{(1,0)}-v^d_{(1,1)}\right)}$ where $\widehat{Pr}$ represents the McFadden kernel smoothed probability simulator. Note that since $\mathbb{1}\left(v^d_{(0,0)} \geq 0\right)$ does not depend on the parameter $\beta$, $\widehat{Pr}$ is a continuous function of $\beta$. In addition, by conditioning on firm entry and stay in the market, we incorporate the observed selection and attrition process into the dynamic choice estimation.

Since $E[\chi_1\chi_2 | entry, stay] = Pr(\chi_{1it} = 1, \chi_{2it} = 1 | entry, stay)$, a natural estimator is to choose parameters $\beta$ to minimize the sum of squared errors between $\chi_1\chi_2$ and $\widehat{Pr}(\chi_1 = 1, \chi_2 = 1 | entry, stay)$. That is,

$$\hat{\beta} = argmin\ Q \equiv \frac{1}{2N}\sum_{i,t}\left[\chi_{1it}\chi_{2it} - \widehat{Pr}(\chi_{1it} = 1, \chi_{2it} = 1 | entry, stay)\right]^2 \quad (3)$$

[18] Note we implicitly assume these drawings are independent of each other.

where $\hat{\beta}$ denotes the parameter estimate; $N$ is the number of observations; and the subscripts $(i, t)$ index firm-year observations. Later in our estimations, we first employ a simulated annealing algorithm to find initial values of the parameter estimate and then utilize the Nelder-Mead algorithm to find a solution to the minimization problem.[19] Besides, we use the bootstrap method to compute the standard errors.

In the estimations, the ***sources of identification*** rely on particular types of variation in the data, a strategy similar to that of Das et al. (2007). The structural relationship between total variable costs and the domestic sales and export revenues allows identification of the preference parameters ($\rho$ and $\tilde{\rho}$). The effect of WTO accession on iceberg trade cost is identified from the variation of the export-domestic-sales ratio from pre- to post-WTO accession (and the heterogeneity in WTO-induced tariff reductions across industries in robustness check). The identification of exit probability $(1-\sigma)$ utilizes time-variation of firms' disappearance in the sample, and similarly the transition probability matrices $(\Sigma)$ are identified from time-variation of state variable $(\frac{\psi}{w+m})$.

The identification of sunk entry and fixed costs of export $(\beta)$ in the dynamic choice modelling largely follows Das et al. (2007), and we extend it to product innovation. For export, the differences in exporting frequencies across firms, which only differ in terms of whether they have exported previously, allow identification of sunk entry costs. Then with the sunk entry costs, the frequency of turning between exporting and non-exporting among firms with positive gross profit streams identifies the fixed costs.

For product innovation, the sunk entry costs are identified from the differences in innovating frequencies among similar firms that only differ in terms of whether they have innovated in the previous period. Then, given the innovation entry cost and profits associated with product innovation, the frequency of turning between innovation and non-innovation identifies the fixed cost of innovation. Besides, as standard in the existing literature, we also impose restrictions on the discount rate $(\delta)$ and distributions of the shocks to aid identification.

Table 1 summarizes the sources of identification. One challenging task in the identification is to separate the static and dynamic effects. For example, given a static incentive to innovate, such as a decrease of trade costs due to WTO accession, firms that have never innovated may not be able to innovate immediately, owing to time needed to procure equipment and hire researchers. Hence, if the data contain only firms that have never innovated before the incentive shock, it may be difficult to separate the static and dynamic effects, as the static effect takes time to occur for these firms. In contrast, firms that have innovated previously can utilize the existing facilities to innovate, and thus can respond to the static incentive shock more quickly. Such variation in the responses allows for separation of the static and dynamic effects.

[19] The estimation is implemented in R, using a *GenSA* package (Xiang, Gubian, Suomela, & Hoeng, 2013) for the simulated annealing algorithm and the in-built *optim* function for the Nelder-Mead algorithm.

Table 1. Identification Table

| Parameter | Economic Meaning | Source of Identification |
|---|---|---|
| $\rho, \tilde{\rho}$ | CES preference | Structural relationship between $TVC$ and $R$ and $\tilde{R}$ |
| $\Delta\tau$ | Change of iceberg trade cost | Variation of $\tilde{R}/R$ from pre- to post-WTO accession [1] |
| $1-\sigma$ | Exit probability | Firm-year disappearance in sample |
| $\Sigma$ | Transition probability matrices | State transitions [2] |
| $f_e^e/f$ ($\beta_3$, $\beta_4$) | Sunk export entry cost | Export persistence |
| $f_e/f$ ($\beta_1$) | Fixed export cost | Export switching |
| $f_n^e/f$ ($\beta_5$, $\beta_6$) | Sunk innovation entry cost | Innovation persistence |
| $f_n/f$ ($\beta_2$) | Fixed innovation cost | Innovation switching |
| $\delta$ | Discount rate | Fixed [3] |

Note: We also assume distributions of the shocks to aid identification; $TVC$, total variable cost; $R$, domestic sales revenue; $\tilde{R}$, export revenue;

[1], we additionally utilize the heterogeneity in WTO-induced tariff reductions across industries as a source of identification in robustness check;

[2], the state transitions contain both learning-by-doing (exporting and innovating) and self-selection effects. Additionally, we also estimate a multinomial logit model with excluded instruments, using the control function approach, to isolate the learning-by-doing effect in the robustness check;

[3], we assume 10% discount rate, and also use 5% and 1% as robustness check.

## 3. Data, Variables and Reduced-form Evidence

The data are a set of air-conditioner manufacturing firms in China, sourced from the National Bureau of Statistics (NBS), China. It covers 184 firms in 2000, 232 firms in 2001, 254 firms in 2002, 237 firms in 2003, 201 firms in 2005, 191 firms in 2006, and 206 firms in 2007, where years 2000 and 2001 are pre-WTO accession periods.[20]

The air-conditioner manufacturing industry presents an interesting case to study in that it is manufacturing-intensive, tradable, innovation-intensive, and exposed to international competition. So, it features sufficient export and product innovation activities. Nevertheless, we expect the economic mechanism we discover to hold beyond the particular industry environment, since the identification does not rely on information that is particular to the air-conditioner manufacturing industry.

The dataset contains information on firms' domestic sales revenue (unit: thousand yuan), export revenue (unit: thousand yuan), value of new products (unit: thousand yuan), intermediate inputs (unit: thousand yuan), total wage (unit: thousand yuan), fixed assets annual net average (unit: thousand yuan), and the number of workers (thousand). We deflate the monetary values by using the producer price index (base year: 2005) obtained from *China Statistical Yearbook 2008*.

[20] Data of year 2004 are not included due to lack of information on product innovation.

Total variable cost (*TVC*) is the sum of total wage and intermediate inputs. We divide *TVC* by the number of workers to compute *wm* ($w + m$). Similarly, dividing the fixed assets annual net average by the number of workers, we can calculate the capital intensity (*kl*). Whether a firm conducts product innovation ($\chi_1$) is a dummy variable that takes a value of one if it reports a positive new product value. Similarly, whether a firm exports ($\chi_2$) is a dummy variable, taking a value of one if it reports positive export revenues.

Table 2 reports the summary statistics. First, we observe substantial variations in the data. For example, the standard deviation for domestic sales revenues is as high as five times of its sample average. We will utilize these variations in the estimations. Second, more than 35% firm-year observations report export revenue, while less than 19% firm-year observations conduct product innovation. In the sample, 10% firm-year observations report both export and conduct product innovation. Third, on the intensive margin of product innovation, the average share of new product value in a firm's total sales revenue is 11.29%, and the export intensive margin is around the same level, with a sample average of 12.12%.

Table 2. Summary Statistics

| Variable | Obs | Mean | Std. dev. | Min | Max |
|---|---|---|---|---|---|
| *R* | 1,505 | 423421.6 | 2127498 | 2.198614 | 3.33E+07 |
| *EX* | 1,505 | 160880.6 | 1032184 | 0 | 2.21E+07 |
| *TW* | 1,505 | 14575.22 | 78529.26 | 6.595843 | 2306854 |
| *M* | 1,505 | 464875.1 | 2312630 | 23 | 4.01E+07 |
| *TL* | 1,505 | 0.612936 | 2.195058 | 0.008 | 41 |
| *VNP* | 1,505 | 251432.1 | 2093122 | 0 | 4.17E+07 |
| *kl* | 1,505 | 114810.3 | 238748.2 | 0 | 3781611 |
| *wm* | 1,505 | 461799.2 | 595295.2 | 4600.29 | 7517665 |
| $\chi_1$ | 1,505 | 0.18804 | | | |
| $\chi_2$ | 1,505 | 0.352824 | | | |

Note: *R*: domestic sales revenues; *EX*: export revenues; *TW*: total wage; *M*: intermediate inputs; *TL*: the number of workers; *VNP*: value of new products; *kl*: fixed assets annual net average per worker; *wm*: *TVC* per worker; $\chi_1$: whether a firm innovates; and $\chi_2$: whether a firm exports. Source: NBS, China, 2000-2007.

As illustrated in the theoretical modelling, we expect to observe a positive correlation between export and innovation and a positive correlation between export/innovation with their previous-period status. To check for these reduced-form predictions, we can nonparametrically estimate these probabilities from the sample as the relative frequency.

Table 3. Probabilities of Innovation and Export, conditional on Previous Period's Experience

| | [1] | | [2] | |
|---|---|---|---|---|
| | $\chi_1 = 0$ | $\chi_1 = 1$ | $\chi_2 = 0$ | $\chi_2 = 1$ |
| $\chi_{-1} = 0$ | 0.888<br>(0.0086) | 0.112<br>(0.0086) | 0.7808<br>(0.0119) | 0.2192<br>(0.0119) |
| $\chi_{-1} = 1$ | 0.1988<br>(0.031) | 0.8012<br>(0.031) | 0.1213<br>(0.0187) | 0.8787<br>(0.0187) |

Note: For [1], the rows correspond to $\chi_{-1} = \chi_{1,-1}$; For [2], the rows correspond to $\chi_{-1} = \chi_{2,-1}$; Figures in the brackets are standard errors.

Table 3 reports the estimates of probabilities, conditional on previous-period's status, which clearly confirm that firms that have innovated/exported in the previous period are more likely to continue innovate/export. Compared with a previously non-exporting firm, a previously exporting firm has a probability of export that is three times higher. Similarly, a previously innovating firm has a probability of innovation that is more than six times higher than that of a previously non-innovating firm.

Table 4. Probabilities of Innovation/Export, conditional on Contemporaneous Export/Innovation

| [1] $Pr(\chi_1 \mid \chi_2)$ | | | [2] $Pr(\chi_2 \mid \chi_1)$ | | |
|---|---|---|---|---|---|
| | $\chi_1 = 0$ | $\chi_1 = 1$ | | $\chi_2 = 0$ | $\chi_2 = 1$ |
| $\chi_2 = 0$ | 0.8706<br>(0.0108) | 0.1294<br>(0.0108) | $\chi_1 = 0$ | 0.6939<br>(0.0132) | 0.3061<br>(0.0132) |
| $\chi_2 = 1$ | 0.7043<br>(0.0198) | 0.2957<br>(0.0198) | $\chi_1 = 1$ | 0.4452<br>(0.0295) | 0.5548<br>(0.0295) |

Note: Figures in the brackets are standard errors.

Table 4 reports the probabilities of innovation/export, conditional on contemporaneous export/innovation. Compared with a non-exporting firm, an exporting firm has a higher probability of conducting product innovation. Similarly, compared with a non-innovating firm, an innovating firm has a higher probability of export. That is, a positive correlation (complementarity) exists between the decisions on whether to innovate and export. The correlation between product innovation and export in the sample is 0.2034.

Since we intend to explore the effects of China's WTO accession on product innovation and export, it is of interest to check whether their trends exhibit systematic differences from pre- to post- WTO accession. Figures 2 and 3 present the time trends of probabilities, average shares and average values of product innovation and export respectively. We can clearly observe that these time trends move upward from pre- to post-WTO accession, suggesting a positive association between firms' product innovation and export and China's WTO accession.

Figure 2. Trends of Probability, Average Share and Average Value of Product Innovation from Pre- to Post- WTO Accession

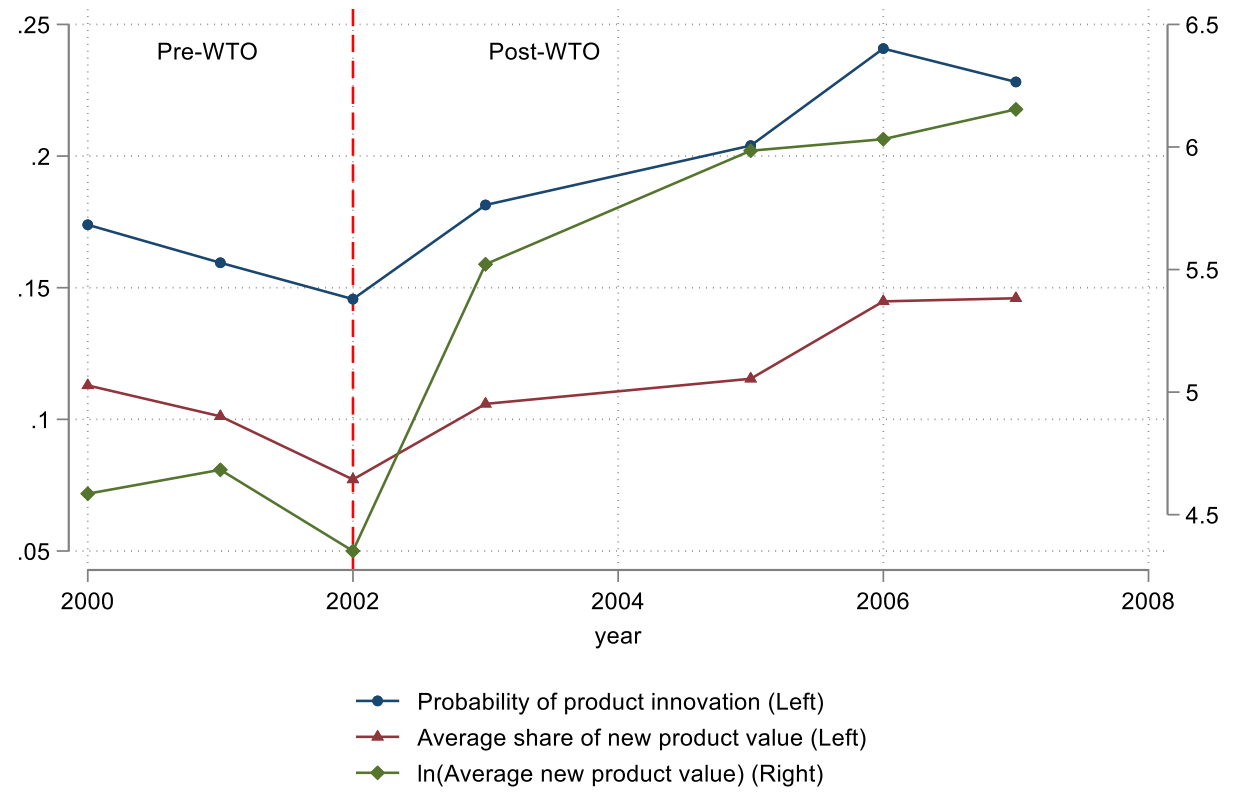

Figure 2. Trends of Probability, Average Share and Average Value of Export from Pre- to Post- WTO Accession

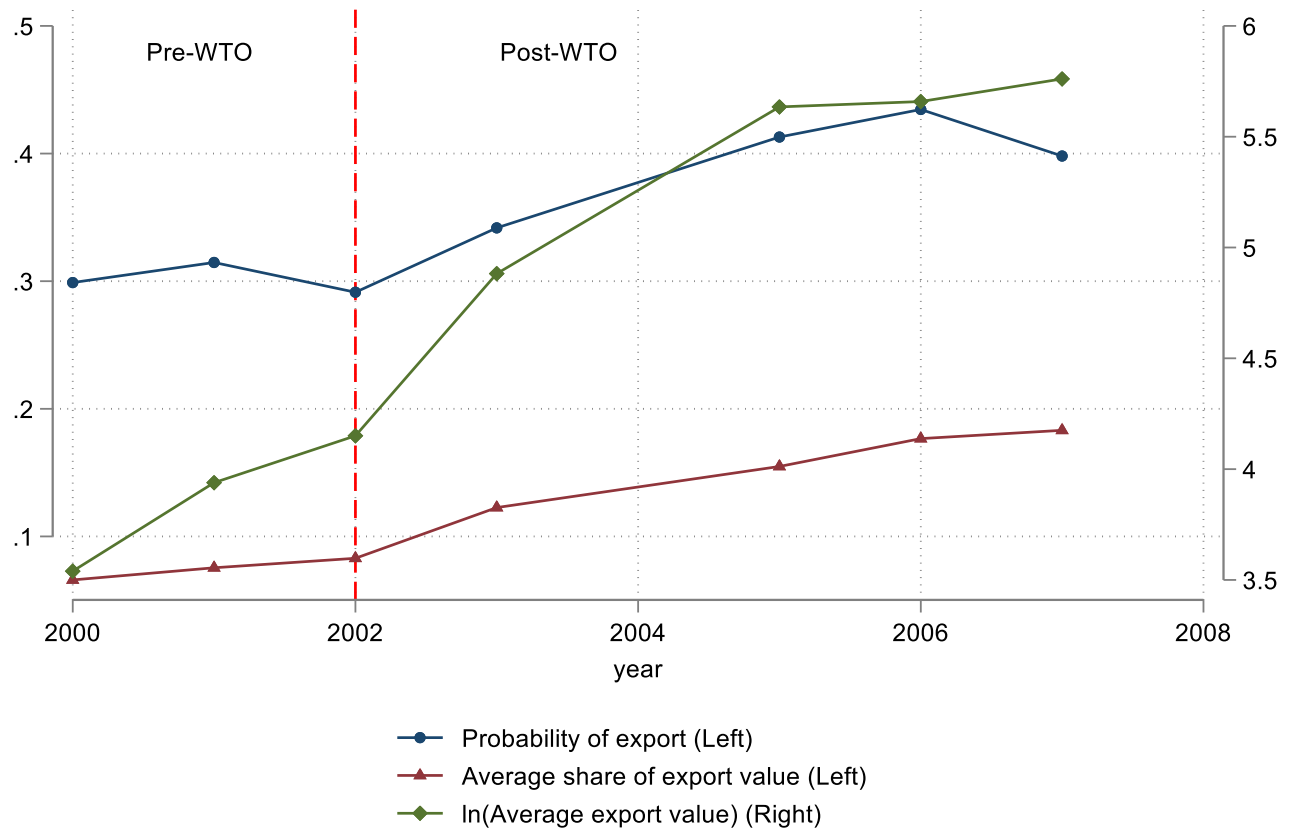

We regress a firm's product innovation status (whether innovate) against different lags of its export status (whether export), in order to examine the persistence of the correlation between export and product innovation. Figure 4 reports the estimates of the correlations, where the dots represent the point estimates and the error bars are 95% confidence intervals. It is clear in Figure 4 that the correlation is positive and persistent across time. The persistence points at dynamic responses of a firm's product innovation to its export participation, and justifies our dynamic modelling of a firms' optimal decisions to export and innovate.

Figure 4. Persistence of the Correlation between Product Innovation Status and the Lags of Export Status

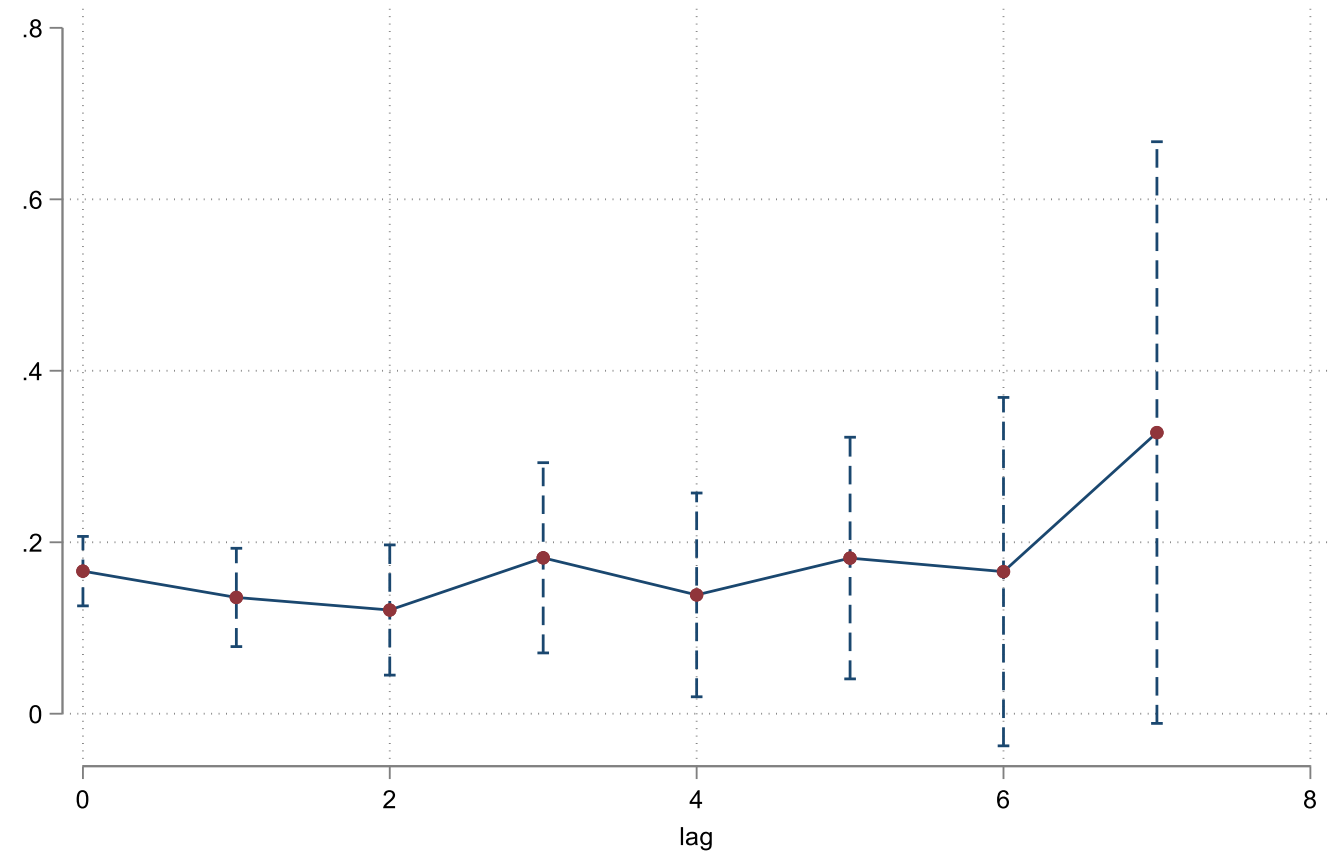

**4. Estimation Results**

We estimate the theoretical model with a sample of air-conditioner manufacturing firms in China from 2000 to 2007, by implementing the algorithm described in section 2.2.

In the first step estimation, we obtain the estimates of the CES preference parameters as $\hat{\rho} = 0.7517$ with a standard error of 0.0051 in the domestic market and $\hat{\tilde{\rho}} = 0.9163$ with a standard error of 0.0079 in the export market, where the hats denote point estimates.[21] The estimates are largely in line with findings in previous studies (see for example Aw et al., 2011; Das et al., 2007; Sun, 2023). Therefore, the elasticity of substitution in the domestic market is 4.03, lower than that in the export market (11.95). The export market consists of the world other than China, which is bigger and likely to have more options of air-conditioners for consumers. Hence, it is not surprising that consumers in the export market have a higher elasticity of substitution.

In step three, we nonparametrically estimate the probability of exit and the transition probability matrices. The exit probability is estimated as the share of firm-year observations where firms appear in this, but not next, year in the total firm-year observations. In the sample, 24.86% firm-year observations do not appear in the next year (standard error 0.0133). Conditional on innovation and export decisions $(\chi_1, \chi_2)$, the transitional probability matrix of $\left(\frac{\psi}{w+m}\right)$ are nonparametrically estimated in a similar way as the relative frequency, where we restrict the sample to firms that appear in both adjacent years (namely conditional on firm entry and stay in the market).

---

[21] A joint estimation of $\rho$ and $\tilde{\rho}$ yields similar results, with $\hat{\rho} = 0.7465$ (standard error 0.0038) and $\hat{\tilde{\rho}} = 0.9245$ (standard error 0.0078). In addition, a test of $\rho = \tilde{\rho}$ rejects the null hypothesis of equality at the 1% level.

*4.1 The effect of WTO accession on iceberg trade cost*

Firms' profit maximization in the domestic and export markets implies a structural relationship between the domestic and export sales revenues, which depends on the CES preference parameters, the aggregate demands and iceberg trade cost. As shown in equation (2), with an appropriate measurement of the aggregate demands and the CES preference parameters, one can link a transformation of domestic sales and export revenues to the iceberg trade cost and a relative price in the base year. Since the base year relative price does not vary across time, one can compare the transformation of domestic sales and export revenues before and after WTO accession to identify the impact of trade liberalization.

Table 5. The Impact of Trade Liberalization on Iceberg Trade Cost

| | [1] | [2] | [3] | [4] |
|---|---|---|---|---|
| $\mathbb{1}(t = 2001)$ | | | -0.0263 | |
| | | | (0.0876) | |
| $\mathbb{1}(t = 2002)$ | | | -0.0559 | |
| | | | (0.0907) | |
| $\mathbb{1}(t = 2003)$ | | | -0.186* | |
| | | | (0.110) | |
| $\mathbb{1}(t = 2005)$ | | | -0.221** | |
| | | | (0.0987) | |
| $\mathbb{1}(t = 2006)$ | | | -0.273*** | |
| | | | (0.100) | |
| $\mathbb{1}(t = 2007)$ | | | -0.266*** | |
| | | | (0.102) | |
| Constant | 1.163*** | 1.170*** | 1.212*** | 1.259*** |
| | (0.153) | (0.156) | (0.158) | (0.149) |
| *dWTO* | -0.135*** | -0.184*** | | -1.527** |
| | (0.0497) | (0.0500) | | (0.747) |
| *dWTO×dT* | | | | 1.424* |
| | | | | (0.758) |
| N | 374 | 374 | 374 | 399 |

Note: In [1], *dWTO* = 1 if post-WTO accession (year > 2001); In [2], *dWTO* = 1 if year > 2002, namely one year lagged; $\mathbb{1}(\cdot)$ is the indicator function; In [4], the implied change of iceberg trade cost from pre- to post-WTO accession is 10.27%. Bootstrap standard errors in parentheses (100 replications); *** p<0.01, ** p<0.05, * p<0.1.

Source: The authors' estimation with data from NBS, China.

The appended error term in equation (2) captures the measurement error and firm level factors. As a national initiative, WTO accession appears unlikely to be correlated with the measurement error and firm level factors in the air-conditioner manufacturing industry. Therefore, we estimate equation (2) by the random effect (RE) estimator. Table 5 reports the estimation results, where in column [1] *dWTO* = 1 if it is after 2001 and in column [2] *dWTO* is one year lagged to allow for the potential lag effect of WTO accession. The estimated coefficients in both columns are negative and statistically significant at the 5% level. Hence, we estimate that the change in the export environment associated with China's WTO accession is equivalent to a 13.5% reduction in iceberg trade costs, which can occur due to a reduction of both tariff and non-tariff barriers. Specifically, compared with pre-WTO accession, the iceberg trade cost reduces by 13.5% after WTO accession (column [1]). The point estimate of the *dWTO* coefficient in column [2] is -0.184, higher than that of column [1]. Nevertheless, one estimate is within one standard deviation of the other.

To what degree do the estimations in columns [1] and [2] pick up the confounding effect of other unobserved factors?[22] For example, one may suspect that technological progress in the transportation sector can reduce the transportation cost, resulting in an estimate of a negative coefficient of *dWTO*. To check for this possibility, we estimate equation (2) with dummy variables for each year, rather than a *dWTO* dummy variable. If the effect picked up by estimate of $\alpha_1$ is due to WTO accession, then we shall observe (1) the estimated coefficients of year dummies are not statistically significant in the pre-WTO periods, and significant in post-WTO periods; and (2) in post-WTO periods, the coefficients of year dummies shall be approximately constant across time.

Column [3] of Table 5 reports the results. One can observe that since 2003, the coefficients of dummy variables are significantly negative, which coincides with China's WTO accession in December of 2001. In addition, the estimates of the coefficients of dummy variables after 2003 are generally within one standard deviation of each other, suggesting the reduction of iceberg trade cost does not significantly vary across years (after 2003). Hence, it appears that the reduction of iceberg trade cost is more driven by WTO accession, rather than other factors such as the technological progress in the transportation sector.[23]

Identification in the regressions of columns [1]-[3] relies on a comparison between pre- and post-WTO accession, namely a difference estimation. In addition, as a robustness check, we also explore a difference-in-differences (DID) identification. In investigating the impact of China's WTO accession on markup dispersion, Lu and Yu (2015) utilize the pre-WTO variation in 2001 tariff rates by industries to achieve a DID identification. Following their spirit, we use the paper pulp manufacturing industry as a control in our DID estimation. The tariff rate in 2001 for the paper pulp manufacturing industry is 0%, while in contrast the tariff rate in 2001 for the air-conditioner manufacturing industry ranges from 15% for parts to as high as 40% for a particular type air-conditioner.[24]

---

[22] In particular, in equation (2), the left-hand-side variable has firm-year variations, while *dWTO* in the right-hand-side only has year variations.

[23] We caution readers that observing the pattern in column [3] does not imply the estimate of $\alpha_1$ picks up the effect of WTO accession. It is a necessary, but not sufficient, condition for our claim.

[24] Tariff data are obtained from https://ttd.wto.org/en/data/idb.

Our DID estimation is a fixed effect estimation with the specification $lny = \alpha_0 + \alpha_1 dWTO + \alpha_2 dWTO \times dT + \varepsilon_2$, where $dT$ is a dummy variable taking a value of 1 if it is the air-conditioner manufacturing industry and 0 if the paper pulp manufacturing industry. Its identification relies on the heterogeneity in WTO-induced tariff reductions across industries. For firms in the air-conditioner manufacturing industry, the change of iceberg trade cost from pre- to post-WTO is equal to $\alpha_1 + \alpha_2$. In constructing $lny$ for firms in the paper pulp manufacturing industry, we estimate $\rho$ and $\tilde{\rho}$ in the same way as discussed above. When testing for equality of $\rho$ and $\tilde{\rho}$, we obtain a $F$-test statistic of 4.78, which fails to reject the null hypothesis of equality at the 1% level. Accordingly, we estimate $\hat{\rho} = \hat{\tilde{\rho}} = 0.7579$ (standard error 0.0033). For the parallel trend testing, we regress $lny$ against $dT$ with pre-WTO sample, and fail to detect a statistically significant coefficient estimate of $dT$ ($t$-statistic 0.75).

Column [4] of Table 5 presents the fixed effect DID estimation results. The coefficient estimates imply a 10.27% reduction of iceberg trade cost after WTO accession for firms in the air-conditioner manufacturing industry. The magnitude is smaller than that of column [1], suggesting that using the paper pulp manufacturing industry as a control dampens the estimated effect of WTO accession. Nevertheless, 10.27% is still within one standard deviation of 13.5% (standard error 0.0497). Hence, our estimate is robust to alternative DID identification strategy.

In addition, we also conduct an event-study type comparison with the paper pulp manufacturing industry by implementing a fixed effect estimation with the specification $lny = \alpha_0 + \sum_{yr \neq 2002} \alpha_{yr} \mathbb{1}(t = yr) \times dT + \varepsilon_2$. Figure A1 in the appendix report the estimation results, which points at a systematic difference from pre- to post-WTO accession, further supporting our DID identification of the effect of WTO accession on iceberg trade cost.

*4.2 The transition probability matrices*

Table 6 reports the non-parametric estimates of the transition probability matrices ($\hat{\Sigma}$) of state variable $\left(\frac{\psi}{w+m}\right)$, conditional on $(\chi_1, \chi_2)$, and Figures 5 and 6 present the corresponding cumulative distribution functions (CDFs). Two patterns of first-order stochastic dominance (FOSD) emerge here.

First, the transition probabilities of export and innovation, $\hat{\Sigma}(\chi_1 = 1, \chi_2 = 1)$, largely first-order stochastically dominate those of no export and no innovation, $\hat{\Sigma}(\chi_1 = 0, \chi_2 = 0)$. That is, firms ex ante predict that export and innovation help them move to a better productivity state in the next period.[25] For example, at state 1, a firm has a probability of 0.2887 moving to a better state in the next period if it does not innovate or export. In contrast, this probability is 0.5 if it exports and innovates.

---

[25] Note the transition probability matrices represent firms' prediction of the law of transition, and only capture the extensive margin of innovation, it is unclear whether product innovation exhibits increasing/decreasing returns. However, Bilir and Morales (2020) find decreasing returns to innovation.

Besides, the FOSD for $\hat{\Sigma}(\chi_1 = 0, \chi_2 = 1)$ and $\hat{\Sigma}(\chi_1 = 1, \chi_2 = 0)$ appears less clear.

Table 6. The Transition Probability Matrices

| | 1 | 2 | 3 | 4 | 1 | 2 | 3 | 4 |
|---|---|---|---|---|---|---|---|---|
| | $(\chi_1 = 0, \chi_2 = 0)$ | | | | $(\chi_1 = 0, \chi_2 = 1)$ | | | |
| 1 | 0.7113 | 0.2535 | 0.0282 | 0.0070 | 0.7 | 0.2667 | 0.0333 | 0 |
| | (0.0845) | (0.0468) | (0.0150) | (0.0075) | (0.1989) | (0.1089) | (0.0356) | (0) |
| 2 | 0.2035 | 0.5752 | 0.1947 | 0.0265 | 0.1136 | 0.5909 | 0.2955 | 0 |
| | (0.0425) | (0.0764) | (0.0415) | (0.0148) | (0.0545) | (0.1404) | (0.0924) | (0) |
| 3 | 0.0194 | 0.1748 | 0.7184 | 0.0874 | 0.0227 | 0.1136 | 0.7045 | 0.1591 |
| | (0.0134) | (0.0417) | (0.0940) | (0.0289) | (0.0208) | (0.0476) | (0.1345) | (0.0570) |
| 4 | 0.0175 | 0.0526 | 0.1579 | 0.7719 | 0 | 0.0202 | 0.0808 | 0.8990 |
| | (0.0177) | (0.0310) | (0.0556) | (0.1453) | (0) | (0.0145) | (0.0291) | (0.1017) |
| | $(\chi_1 = 1, \chi_2 = 0)$ | | | | $(\chi_1 = 1, \chi_2 = 1)$ | | | |
| 1 | 0.6429 | 0.3571 | 0 | 0 | 0.5 | 0.25 | 0 | 0.25 |
| | (0.2923) | (0.2020) | (0) | (0) | (0.4943) | (0.3205) | (0) | (0.3205) |
| 2 | 0 | 0.6190 | 0.3333 | 0.0476 | 0 | 0.6667 | 0.3333 | 0 |
| | (0) | (0.2007) | (0.1392) | (0.0494) | (0) | (0.3589) | (0.2303) | (0) |
| 3 | 0.0417 | 0.0833 | 0.6667 | 0.2083 | 0.0323 | 0.0323 | 0.8065 | 0.1290 |
| | (0.0403) | (0.0574) | (0.1770) | (0.0926) | (0.0335) | (0.0335) | (0.1883) | (0.0682) |
| 4 | 0.0526 | 0 | 0.1579 | 0.7895 | 0 | 0 | 0.0227 | 0.9773 |
| | (0.0507) | (0) | (0.0898) | (0.2265) | (0) | (0) | (0.0217) | (0.1362) |

Note: Transition is from rows to columns; For example, with $(\chi_1 = 0, \chi_2 = 0)$, a firm at state 1 has a probability of 0.2535 to transition to state 2 in the next period; Figures in brackets are standard errors, and note that when a probability is estimated to be 0/1, its S.E. is 0.

Source: The authors' estimation with data from NBS, China.

Figure 5. First Order Stochastic Dominance by Innovation and Export

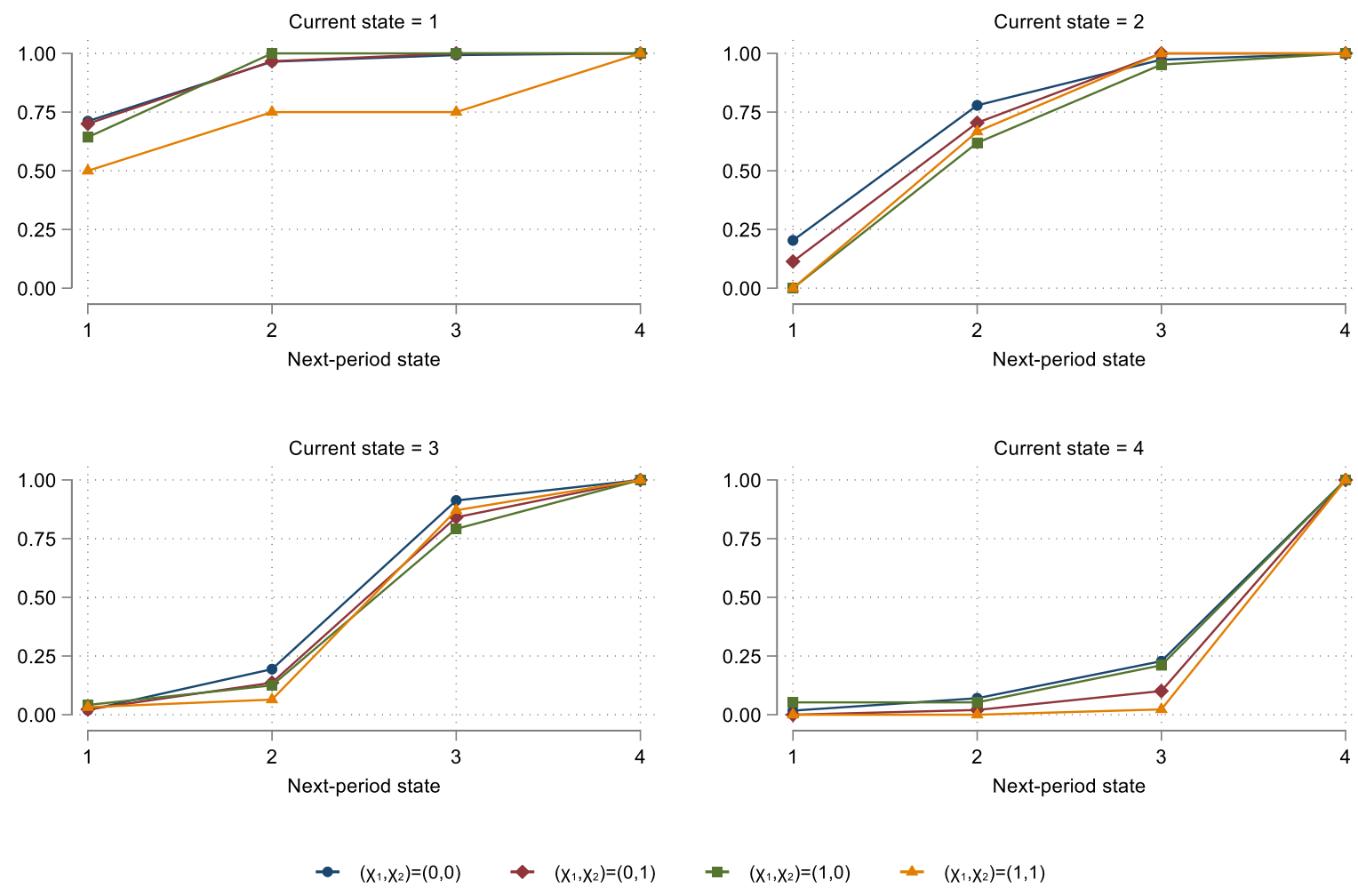

Second, the transition probabilities at a higher productivity state also largely first-order stochastically dominate those at a lower productivity state. For example, for $(\chi_1 = 0, \chi_2 = 0)$, the CDF at state 1 lies above that of state 4. One exception is the transition probability matrix of $(\chi_1 = 1, \chi_2 = 1)$. At state 1, a firm has a probability of 0.5 to stay in the same sate next period, and has 0.25 probability of moving to states 2 and 4 respectively. This results in a CDF below the CDFs of states 2 and 3 at state 3 (bottom right panel of Figure 6), violating FOSD. This occurs possibly due to limited sample size, since only a small proportion of firms both export and innovate.

As per Proposition 2, the first-order stochastic dominance leads to an additional dynamic benefit of export and product innovation. That is, as a firm exports and innovates, it predicts to move to a better productivity state in the future, and in addition moving to a better productivity state further improves the chance of upward movement in the next period.

Figure 6. First Order Stochastic Dominance by States

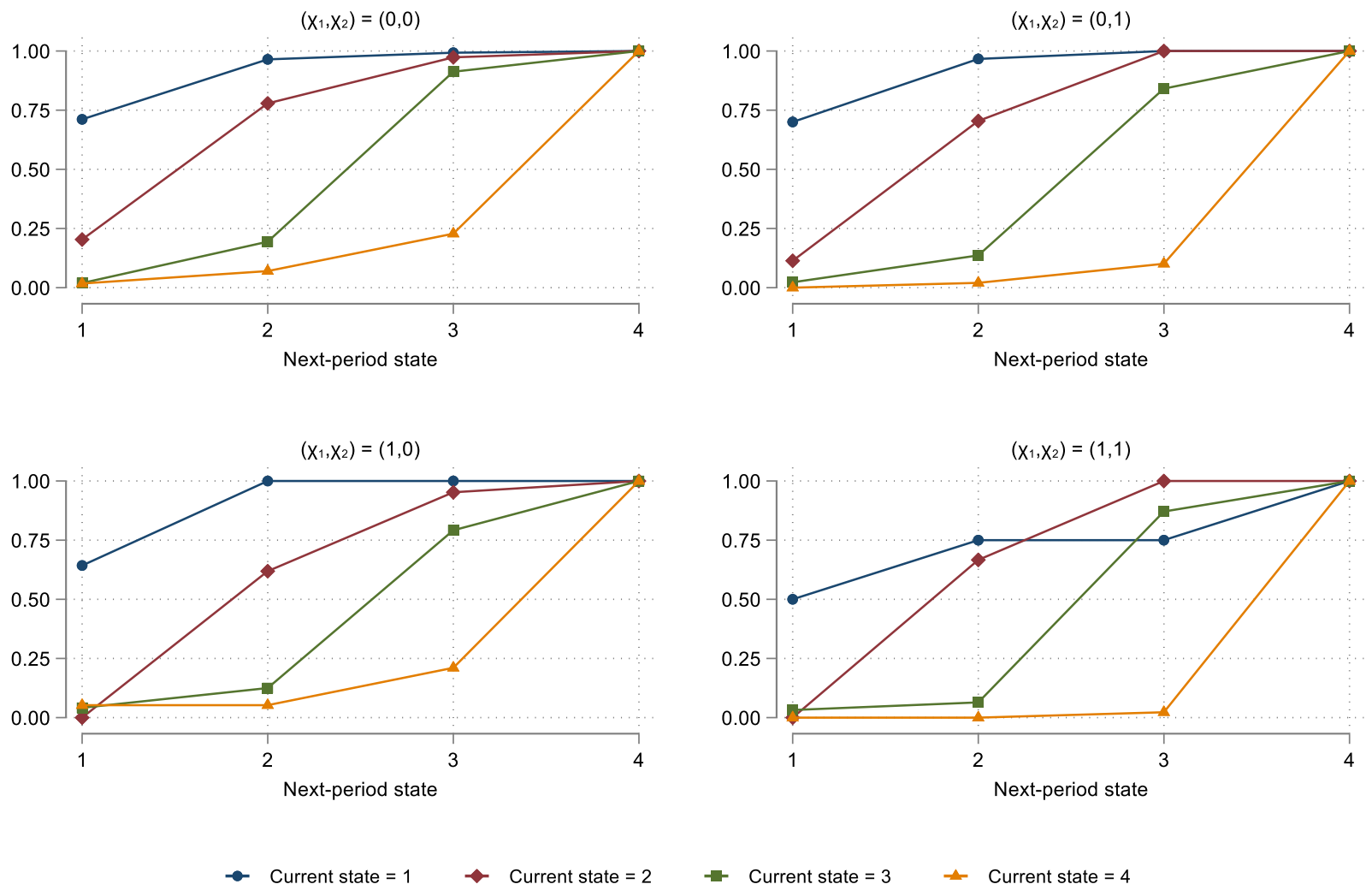

*4.3 The dynamic decisions on export and product innovation*

As discussed in section 2.2, we implement a nested fixed-point algorithm in estimating the dynamic choices of export and product innovation. We utilize the post-WTO data in the estimation, leaving the pre-WTO data for counterfactual analysis.

Table 7 reports the estimation results, where the upper panel presents results with 10% discount rate, on which our interpretation will be based, while the middle and lower panels are the results with 5% and 1% discount rate respectively, which serve as a robustness check. The parameter $\beta_0$, which contains the post-WTO iceberg trade cost $(\tau^{\frac{\tilde{\rho}}{\tilde{\rho}-1}} f^{\frac{(1-\rho)\tilde{\rho}}{\rho(1-\tilde{\rho})}} \tilde{A} A^{-\frac{(1-\rho)\tilde{\rho}}{\rho(1-\tilde{\rho})}} \equiv e^{\beta_0})$, is estimated to be close to -8.97 (namely $e^{\beta_0} = 0.000127$), statistically significant at the 5% level. Hence, with a 13.5% decrease of iceberg trade cost due to the WTO accession, the pre-WTO $\beta_0$ is -10.4484.

Regarding the fixed costs of export ($\frac{f_e}{f} \equiv e^{\beta_1} + \varepsilon_3$), the point estimate of $\beta_1$ is positive, but not statistically significant at the 5% level. The point estimate of $\beta_2$ is negative and significant at the 5% level, suggesting a relatively small fixed cost of innovation ($\frac{f_n}{f} \equiv e^{\beta_2} + \varepsilon_4$). In contrast, the entry costs of export and innovation appear to be large. For the export entry cost ($\frac{f_e^e}{f} \equiv e^{\beta_3+\beta_4 IsBig} + \varepsilon_5$), the point estimates of $\beta_3$ and $\beta_4$ are 8.3173 and -2.3615, respectively, both of which are significant at the 5% level. Similarly, the point estimates of $\beta_5$ and $\beta_6$ ($\frac{f_e^n}{f} \equiv e^{\beta_5+\beta_6 IsBig} + \varepsilon_6$) are 8.4302 and -1.404, respectively and significant at the 5% level.

Table 7. Estimation Results of the Dynamic Choices of Export and Product Innovation

| | | Coef | S.E. | t |
|---|---|---|---|---|
| [1] Discount rate = 10% | | | | |
| Constant | $\beta_0$ | -8.9705*** | 0.1393 | -64.3761 |
| Fixed cost of export | $\beta_1$ | 0.4345 | 0.4417 | 0.9838 |
| Fixed cost of innovation | $\beta_2$ | -1.7658*** | 0.5318 | -3.3203 |
| Entry cost of export | $\beta_3$ | 8.3171*** | 0.2633 | 31.5883 |
| Entry cost of export, big firm | $\beta_4$ | -2.3615*** | 0.4415 | -5.3488 |
| Entry cost of innovation | $\beta_5$ | 8.4302*** | 0.2438 | 34.5849 |
| Entry cost of innovation, big firm | $\beta_6$ | -1.404*** | 0.3791 | -3.7037 |
| [2] Discount rate = 5% | | | | |
| Constant | $\beta_0$ | -8.4252*** | 0.1679 | -50.1775 |
| Fixed cost of export | $\beta_1$ | 0.9274*** | 0.4171 | 2.2233 |
| Fixed cost of innovation | $\beta_2$ | -4.8145*** | 0.3658 | -13.1598 |
| Entry cost of export | $\beta_3$ | 8.7585*** | 0.2701 | 32.4267 |
| Entry cost of export, big firm | $\beta_4$ | -1.9798*** | 0.4512 | -4.3875 |
| Entry cost of innovation | $\beta_5$ | 9.1583*** | 0.2235 | 40.9749 |
| Entry cost of innovation, big firm | $\beta_6$ | -1.4650*** | 0.3830 | -3.8254 |
| [3] Discount rate = 1% | | | | |
| Constant | $\beta_0$ | -7.9947*** | 0.1421 | -56.2516 |
| Fixed cost of export | $\beta_1$ | 1.3310*** | 0.4125 | 3.2265 |
| Fixed cost of innovation | $\beta_2$ | -3.3665*** | 0.3876 | -8.6851 |
| Entry cost of export | $\beta_3$ | 9.0410*** | 0.3140 | 28.7902 |
| Entry cost of export, big firm | $\beta_4$ | -1.7304*** | 0.4896 | -3.5344 |
| Entry cost of innovation | $\beta_5$ | 9.6435*** | 0.1757 | 54.8824 |
| Entry cost of innovation, big firm | $\beta_6$ | -1.6390*** | 0.3964 | -4.1344 |

Note: Standard errors are bootstrapped (50 replications); *** p<0.01; Fixed and entry costs are relative to fixed cost of production; N = 1089.

Source: The authors' estimation with data from NBS, China.

Comparing the estimation with 10% discount rate with those of 5% and 1% discount rates, we observe variations in point estimates. However, the pattern of small fixed costs and large entry costs continues to hold. Such a pattern corroborates with the finding of Das et al. (2007) on exporting, while our results here extend to innovation. The large entry costs of export and innovation confirm a dynamic benefit from trade liberalization as per Proposition 2, namely as a firm innovates (and exports) in the current period, they can save the entry costs, which are substantial, in the future.

To assess the model fit, let $z \equiv \left(\frac{\psi}{w+m}, \chi_{1,-1}, \chi_{2,-1}, IsBig\right)$ and $Z \equiv \{z\}$, where $\frac{\psi}{w+m}$ is the discretised states and $|Z| = 32$. On the one hand, for each $z$, we can simulate the probability of export, probability of product innovation, and probability of both export and product innovation, given the parameter estimates. With the relative frequency for each $z$ computed from data, we then compute the simulated probabilities of export, product innovation, and both (Column [Model] in Table 8). On the other hand, we can non-parametrically compute these probabilities from data (Column [Data] in Table 8). Consequently, the difference between these two sets of probabilities (Column [Difference] in Table 8) informs the model fit. Table 7 presents the results. We can observe that the differences for probabilities of product innovation and both export and product innovation are small. In contrast, the model underpredicts the probability of export, compared with data. This occurs possibly due to the fact that a sizeable proportion of firms are engaged in processing trade. Later we will conduct a robustness check by restricting the sample to only non-processing trade firms.

Table 8. Model Fit

| Moments | Data | Model | Difference |
|---|---|---|---|
| $Pr(\chi_1 = 1)$ | 0.1965 | 0.1984 | 0.0019 |
| $Pr(\chi_2 = 1)$ | 0.3701 | 0.2543 | -0.1157 |
| $Pr(\chi_1 = 1, \chi_2 = 1)$ | 0.1084 | 0.1386 | 0.0303 |

Source: The author's estimation using NBS data.

## 5. Counterfactual analysis

*5.1 The effect of trade liberalization on probability of innovation and export*

With the estimates above, we can then conduct counterfactual analysis to assess the effects of trade liberalization (WTO accession) on firms' probability to export and innovate. First, from the data, we can nonparametrically calculate this probability as the relative frequency of firms that both export and innovate in each year, namely $\widetilde{\Pr}(\chi_{1t} = 1, \chi_{2t} = 1) = \frac{1}{N_t}\sum_{i=1}^{N_t} \mathbb{1}(\chi_{1it} = 1, \chi_{2it} = 1)$ where $\widetilde{\Pr}$ denotes the probability estimate; and $N_t$ is the number of firms in year *t*.

Second, using the estimated parameters, we can simulate $\widetilde{\Pr}(\chi_{1t} = 1, \chi_{2t} = 1|z)$ where $\widetilde{\Pr}$ represents the simulated probability. Then $\widetilde{\Pr}(\chi_{1t} = 1, \chi_{2t} = 1) = \sum_{z_t \in Z} \widetilde{\Pr}(\chi_{1t} = 1, \chi_{2t} = 1|z)\widehat{\Pr}(z_t)$, where $\widehat{\Pr}(z_t)$ is computed nonparametrically from data as the relative frequency in each year. The probability $\widetilde{\Pr}(\chi_{1t} = 1, \chi_{2t} = 1)$ serves as the counterfactual. We exclude the samples of years 2000 and 2005 due to missing $(\chi_{1,-1}, \chi_{2,-1})$, and the post-WTO is after 2002 (one year after the WTO accession) to allow for lagged effect from WTO accession.

For post-WTO accession, the difference between $\widehat{\Pr}(\chi_{1t} = 1, \chi_{2t} = 1)$ and $\widetilde{\Pr}(\chi_{1t} = 1, \chi_{2t} = 1)$ is due to estimation errors. For pre-WTO accession, this difference is due to both estimation errors and a higher iceberg trade cost. Hence, we can infer the effect of WTO accession on these probabilities in a difference-in-differences manner.

Figure 7. The Probabilities of Product Innovation and Export

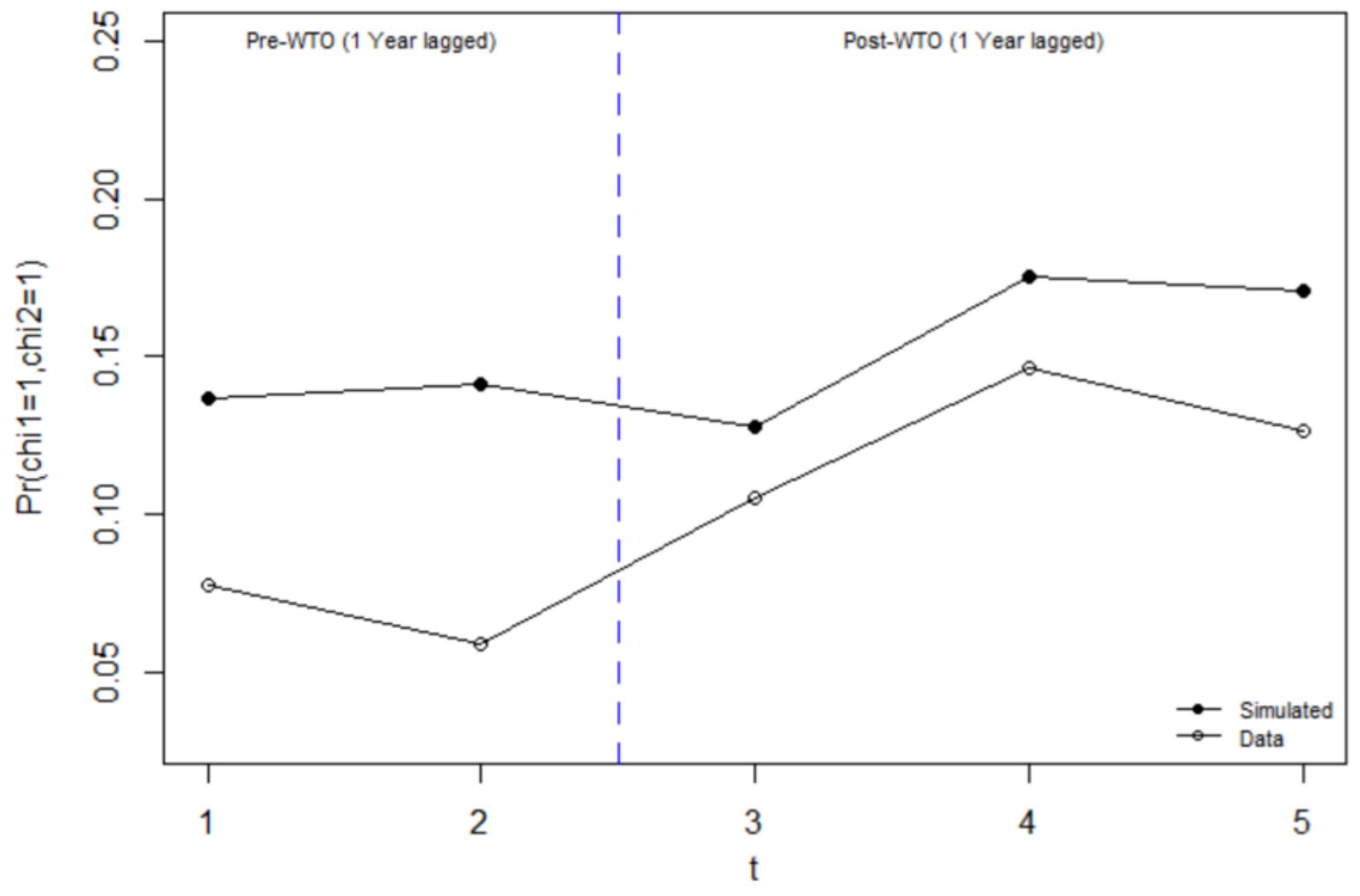

Note: $t = 1$, year 2001; $t = 2$, year 2002; $t = 3$, year 2003; $t = 4$, year 2006; $t = 5$, year 2007.

Figure 7 presents the results. In all five years, the simulated probabilities of export and product innovation appear to be higher than their non-parametrically computed counterparts. However, the difference after WTO accession is smaller than that pre-WTO accession. Then the difference between these two differences is due to WTO accession that lowers the iceberg trade cost. Should there be a lower iceberg trade cost pre-WTO, the probability of export and product innovation will be higher than overserved in the data. That is, the WTO accession indeed promotes firms' probability of export and product innovation. A difference-in-differences regression finds that WTO accession promotes the probability of export and product innovation by 0.0386.

*5.2 The relative contribution of three channels*

Trade liberalization, namely China's WTO accession, reduces the iceberg trade cost by around 13.5%, which promotes firms export and subsequently product innovation, via one static and two dynamic channels. Nevertheless, how important are these two dynamic channels? We explore this issue by simulating $Pr(\chi_1 = 1, \chi_2 = 1 | \chi_{1,-1} = 1, \chi_{2,-1} = 1, entry, stay)$ and comparing the simulated probabilities in different scenarios, namely pre-WTO, post-WTO, shutting down the channel of transitioning into a better state, and shutting down the channel of entry cost saving. In dosing so, we fix the transition probability matrices to be the identity matrices other than when a firm decides to both innovate and export, which is set as follows:

$$\begin{pmatrix} 1 & 0 & 0 & 0 \\ 0 & 0.6 & 0.4 & 0 \\ 0 & 0 & 0.4 & 0.6 \\ 0 & 0 & 0 & 1 \end{pmatrix}.$$

That is, when a firm decides to both innovate and export, it has a chance of moving to a better state in the next period in states 2 and 3. Otherwise, its state will stay the same.

Let $\mathcal{A}$, $\mathcal{B}$, $\mathcal{C}$, $Pre$ and $Post$ denote the events of static effect, state transition, entry cost saving, pre- and post-WTO accession respectively, and we assume the following holds:

(1) $\mathcal{A}$, $\mathcal{B}$, $\mathcal{C}$, $Pre$ and $Post$ are independent of each other and independent of the event $\{\chi_{1,-1} = 1, \chi_{2,-1} = 1, entry, stay\}$;

(2) $Pr(Pre)$ and $Pr(Post)$ do not depend on starting state, and $Pr(\mathcal{A})$, $Pr(\mathcal{B})$, and $Pr(\mathcal{C})$ may depend on starting state;

(3) $Pr(\mathcal{A}) \in (0,1)$, $Pr(\mathcal{B}) \in (0,1)$, $Pr(\mathcal{C}) \in (0,1)$, and $Pr(\mathcal{A} \cap \mathcal{B} \cap \mathcal{C}) \neq 0$.

The independence in (1) facilitates the computation, and we can relax to only require $Pre$ and $Post$ to be independent of the other events. The event of WTO accession is a national initiative, and it appears plausible that it is independent of firm-level events. In (2), given that WTO accession is a national-level event, it is unlikely that it will depend on firms' starting state, while in contrast, firm-level events can depend on firm's starting states. Assumption (3) is to avoid conditioning on events of zero probability.

Then we can decompose the effect of trade liberalization on the probability of product innovation and export, conditional on $\{\chi_{1,-1} = 1, \chi_{2,-1} = 1, entry, stay\}$, as follows:

$$\underbrace{Pr(\chi_1 = 1, \chi_2 = 1 | Post, \mathcal{A}, \mathcal{B}, \mathcal{C}) - Pr(\chi_1 = 1, \chi_2 = 1 | Pre, \mathcal{A}, \mathcal{B}, \mathcal{C})}_{Total\ effect\ of\ trade\ liberalization}$$

$$= \kappa \left\{ \left[ \underbrace{Pr(\chi_1 = 1, \chi_2 = 1 | Post, \mathcal{A}, \mathcal{B}, \mathcal{C}^c) - Pr(\chi_1 = 1, \chi_2 = 1 | Pre, \mathcal{A}, \mathcal{B}, \mathcal{C}^c)}_{No\ transition} \right] \right.$$

$$+ \left[ \underbrace{Pr(\chi_1 = 1, \chi_2 = 1 | Post, \mathcal{A}, \mathcal{B}^c, \mathcal{C}) - Pr(\chi_1 = 1, \chi_2 = 1 | Pre, \mathcal{A}, \mathcal{B}^c, \mathcal{C})}_{No\ entry\ cost\ saving} \right]$$

$$\left. + \left[ \underbrace{Pr(\chi_1 = 1, \chi_2 = 1 | Post, \mathcal{A}, \mathcal{B}^c, \mathcal{C}^c) - Pr(\chi_1 = 1, \chi_2 = 1 | Pre, \mathcal{A}, \mathcal{B}^c, \mathcal{C}^c)}_{No\ transition\ +\ No\ entry\ cost\ saving} \right] \right\}$$

where for notational convenience, we omit conditioning on $\{\chi_{1,-1} = 1, \chi_{2,-1} = 1, entry, stay\}$, and $\kappa \equiv \frac{Pr(\mathcal{B}^c)Pr(\mathcal{C}^c)}{Pr(\mathcal{B}^c)Pr(\mathcal{C})+Pr(\mathcal{B})Pr(\mathcal{C}^c)-Pr(\mathcal{B})Pr(\mathcal{C})}$ which does not depend on WTO accession and may depend on firms' starting state. We can simulate these probabilities, compute $\kappa$, and then calculate the relative contribution (share) of these three channels.[26]

Table 9 reports the simulation results. Five patterns emerge. First, the prediction of transition to a better productivity state and entry cost saving play important roles in determining the probability of export and product innovation. For example, in pre-WTO at state 2, the probability of export and product innovation is 0.9368, given the firm exports and innovates in the previous period. It drops to 0.1569 once the channel of transition to a better productivity state is shut down, which further drops to 0.1134 if, additionally, the entry cost saving channel is shut down. Second, from pre- to post-WTO, namely a reduction of iceberg trade cost, we observe the probabilities weakly increase at each state and either with or without transition and entry cost saving channels, shown by the all non-negative figures in the rows of [Effect] in Table 9.

Third, the prediction of transitioning to a better state appears to play an important role in the change of probabilities from pre- to post-WTO. For example, when shutting down the transition channel, the probability increases by 0.0007, 0.0925, 0.1574, and 0.1066 for starting states 1, 2, 3, and 4 respectively. Fourth, similarly, the channel of entry cost saving also plays an important role. Fifth, using the above decomposition, we quantify the relative contributions of three channels, reported in rows [Share] in Table 9. It can be observed that the static channel (column [No transition + no entry cost saving] in Table 9) only plays a modest (though pivotal) role, while the roles of dynamic channels are more significant. For example, at the starting state 3, the channel of entry cost saving with static effect accounts for 62.38% of the total effects of WTO accession (row [3]-[Share], column [No transition] in Table 9). The channel of transition with static effect accounts for 28.46% of the total effects (row [3]-[Share], column [No entry cost saving] in Table 9). In contrast, the share of only static channel is 9.16% (row [3]-[Share], column [No transition + no entry cost saving] in Table 9).

[26] Note we cannot shut down the static channel since it jump-starts the subsequent dynamic channels. That is, shutting down the static channel will result in no change of probability of innovation and export from pre- to post-WTO, irrespective of the dynamic channels.

Table 9. The Relative Contributions of Three Channels

| | | All three channels | No transition | No entry cost saving | No transition + no entry cost saving |
|---|---|---|---|---|---|
| [1] | Pre | 0.0859 | 0.0859 | 0.0000 | 0.0000 |
| | Post | 0.0866 | 0.0866 | 0.0000 | 0.0000 |
| | Effect | 0.0007 | 0.0007 | 0.0000 | 0.0000 |
| | Share | 100.00% | 100.00% | 0.00% | 0.00% |
| [2] | Pre | 0.9368 | 0.1569 | 0.1134 | 0.0000 |
| | Post | 0.9601 | 0.2494 | 0.2509 | 0.0000 |
| | Effect | 0.0233 | 0.0925 | 0.1375 | 0.0000 |
| | Share | 100.00% | 40.22% | 59.78% | 0.00% |
| [3] | Pre | 0.8835 | 0.4463 | 0.0645 | 0.0085 |
| | Post | 0.9340 | 0.6036 | 0.1363 | 0.0316 |
| | Effects | 0.0505 | 0.1574 | 0.0718 | 0.0231 |
| | Share | 100.00% | 62.38% | 28.46% | 9.16% |
| [4] | Pre | 0.7593 | 0.7593 | 0.0732 | 0.0732 |
| | Post | 0.8659 | 0.8659 | 0.1485 | 0.1485 |
| | Effect | 0.1066 | 0.1066 | 0.0753 | 0.0753 |
| | Share | 100.00% | 41.46% | 29.27% | 29.27% |

Note: Figures in rows [Pre] and [Post] are the simulated $Pr(\chi_1 = 1, \chi_2 = 1 | \chi_{1,-1} = 1, \chi_{2,-1} = 1, entry, stay)$; Effect = Post - Pre, which captures the total effect of trade liberalization; Rows [Share] present the relative contribution of the three channels from the decomposition; [1], [2], [3], and [4] correspond to starting states 1, 2, 3, and 4 respectively.

## 6. Robustness Checks

In section 4, we have conducted two robustness checks, namely a DID identification of the effect of WTO accession on iceberg trade cost (column 4 of Table 5) and alternative discount rates (middle and bottom panels of Table 7). Our findings continue to hold in these checks. In this section, we undertake four additional robustness checks. First, we attempt to isolate the causal effect of learning by doing (exporting and innovating) in state transition. Second, we check an alternative discretization ($K = 2$). Third, we use total factor productivity to measure states. Fourth, we exclude processing trade firms in the dynamic choice estimation.

*6.1 Sophisticated firms*

In estimating the transition probability matrices, we assume firms predict these probabilities from data. While it seems plausible that firms ex ante gauge the evolution of productivity states by observing what happens to other firms in the industry, this prediction analysis is not ex post causal. That is, the predicted probabilities contain both the ex post causal effect of export and innovation, namely learning by doing (exporting and innovating), and the self-selection effect, namely more productive firms self-select to export and innovate. If firms are sophisticated, they may recognize the confounding effect of self-selection. In such a case, will the findings on dynamic decisions of export and innovation change? In this subsection, we explore this possibility as a robustness check.

To do so, we first parameterize the transition probabilities as a multinomial logit (MNL) model, as follows:

$$Pr\left(\left(\frac{\psi}{w+m}\right)_+ = k\right) = \frac{e^{\lambda_0^k+\lambda_1^k\chi_1+\lambda_2^k\chi_2+\sum_{j=1}^{K-1}\lambda_{2+j}^k d\psi_j}}{1+\sum_{\tilde{k}=1}^{K-1} e^{\lambda_0^{\tilde{k}}+\lambda_1^{\tilde{k}}\chi_1+\lambda_2^{\tilde{k}}\chi_2+\sum_{j=1}^{K-1}\lambda_{2+j}^{\tilde{k}} d\psi_j}}$$

where $\frac{\psi}{w+m} \in \{1,2,\dots,K-1\}$ is the discretized productivity state, state $K$ is chosen as the base outcome, and the subscript $+$ denotes next period; $d\psi_j$ is a dummy variable, taking a value of 1 if the current state $\frac{\psi}{w+m} = j$ and 0 otherwise; the $\lambda$s are parameters to be estimated from data. Due to self-selection, $\chi_1$ (product innovation) and $\chi_2$ (export) are potentially endogenous. We then use the control function approach to address the potential endogeneity issue. That is, using the shares of number of firms in the province-year that innovates and exports (excluding a firm itself) and their average innovation and export intensity as excluded instruments, we first estimate a bivariate probit model, where the dependent variables are $\chi_1$ and $\chi_2$ and the independent variables are the excluded instruments and $d\psi_j$. We then predict the probabilities and compute generalized error terms. The error terms are then included in the second-stage MNL maximum likelihood estimation. To account for the fact the generalized error terms are estimated from the first-stage regression, we compute the standard error by bootstrapping. In addition, we can test for the joint significance of the generalized error terms in the second-stage MNL estimation to check the presence of endogeneity.

The shares of firms in the province-year that innovate and export (excluding a firm itself) and their average innovation and export intensity appear to be valid excluded instruments. On the one hand, they capture market-wide shocks which also affect a firm's optimal innovation and export decisions, and hence are correlated with $\chi_1$ and $\chi_2$. On the other hand, conceptually, they shall not affect the evolution of a firm's productivity state directly.

Table A1 in the appendix reports the estimation results, including the first-stage bivariate probit model, the second-stage MNL model, and a MNL model assuming $\chi_1$ and $\chi_2$ being exogenous. In testing for joint significance of the generalized error terms, we obtain a chi-squared test statistic of 7.55 with a $p$-value 0.2733. Hence, at the 5% level, we fail to reject the null hypothesis of no endogeneity (no self-selection). However, we still use the point estimates of coefficients in the second-stage MNL to predict the transition probabilities, and Figures A2 and A3 in the appendix report their CDFs, where the FOSD continues to hold.

Table 10. Estimation Results with Sophisticated Firms

| | | Coef | S.E. | t |
|---|---|---|---|---|
| Constant | $\beta_0$ | -9.8817*** | 0.2548 | -38.7826 |
| Fixed cost of export | $\beta_1$ | -0.0908 | 0.6559 | -0.1385 |
| Fixed cost of innovation | $\beta_2$ | -1.7177*** | 0.6361 | -2.7006 |
| Entry cost of export | $\beta_3$ | 7.7139*** | 0.3820 | 20.1951 |
| Entry cost of export, big firm | $\beta_4$ | -1.8037*** | 0.5254 | -3.4329 |
| Entry cost of innovation | $\beta_5$ | 9.2012*** | 0.3063 | 30.0409 |
| Entry cost of innovation, big firm | $\beta_6$ | -2.4224*** | 0.3579 | -6.7683 |

Note: Standard errors are bootstrapped (50 replications); *** p<0.01; Fixed and entry costs are relative to fixed cost of production; N = 1089.

Source: The authors' estimation with data from NBS, China.

With the predicted transition probabilities, we then re-estimate the dynamic decisions to innovate and export, and Table 10 reports the estimation results. Comparing Tables 10 with 7, we can observe variations in the point estimates of coefficients. Nevertheless, the pattern of large entry costs and small fixed costs continue to hold. Hence, in this sense, our findings are robust.

*6.2 Two-state estimation*

An empirical pattern of export and innovation is that only a small proportion of firms exports or innovates, and an even smaller proportion of firms does both. Consequently, in estimating the transition probability matrices, particularly that of $(\chi_1, \chi_2) = (1,1)$, an estimate of zero probability can be due to either the true transition probability being zero or lack of data. To check whether our findings are robust here, we also conduct estimations with two states. Table 11 reports the estimates of two-state transition probability matrices, and Table 12 presents the corresponding estimation results of the dynamic choice model of export and product innovation. Comparing Tables 11 and 12 with Tables 6 and 7, first, not surprisingly, we can observe differences in point estimates. Second, our findings continue to hold with two states. That is, firms predict that export and innovation help them move to a better state in the next period, and the entry costs of both export and innovation are large, while fixed costs are small. Hence, our findings appear to be robust to different levels of states.

Table 11. Two-state Transition Probability Matrices

| | 1 | 2 | 1 | 2 |
|---|---|---|---|---|
| | $(\chi_1 = 0, \chi_2 = 0)$ | | $(\chi_1 = 0, \chi_2 = 1)$ | |
| 1 | 0.8057 | 0.1943 | 0.7241 | 0.2759 |
| | (0.0271) | (0.0271) | (0.0648) | (0.0648) |
| 2 | 0.1324 | 0.8676 | 0.0503 | 0.9497 |
| | (0.023) | (0.023) | (0.0188) | (0.0188) |
| | $(\chi_1 = 1, \chi_2 = 0)$ | | $(\chi_1 = 1, \chi_2 = 1)$ | |
| 1 | 0.7037 | 0.2963 | 0.7143 | 0.2857 |
| | (0.0796) | (0.0796) | (0.1569) | (0.1569) |
| 2 | 0.098 | 0.902 | 0.037 | 0.963 |
| | (0.0432) | (0.0432) | (0.025) | (0.025) |

Source: The authors' estimation with data from NBS, China.

Table 12. Two-state Estimation Results

| | | Coef. | S.E. | t |
|---|---|---|---|---|
| Constant | $\beta_0$ | -2.6903*** | 0.1204 | -22.3499 |
| Fixed cost of export | $\beta_1$ | 0.9372 | 0.5247 | 1.7860 |
| Fixed cost of innovation | $\beta_2$ | -9.2793*** | 0.6332 | -14.6552 |
| Entry cost of export | $\beta_3$ | 8.7452*** | 0.3166 | 27.6247 |
| Entry cost of export, big firm | $\beta_4$ | -2.2956*** | 0.6308 | -3.6394 |
| Entry cost of innovation | $\beta_5$ | 9.9927*** | 0.1561 | 64.0075 |
| Entry cost of innovation, big firm | $\beta_6$ | -1.1741 | 0.6149 | -1.9094 |

Note: Standard errors are bootstrapped (50 replications); *** p<0.01; Fixed and entry costs are relative to fixed cost of production; N = 1089.

Source: The authors' estimation with data from NBS, China.

### *6.3 Alternative measurement of states*

For the state variable $\left(\frac{\psi}{w+m}\right)$, $\psi$ is measured by the principal component of firm size (total number of workers) and capital intensity (fixed assets per worker). As a robustness check, we also measure $\psi$ as TFP. That is, first, we construct capital stock by the perpetual inventory method, namely $KS_t = 0.95 \times KS_{t-1} + I_t$ where $KS$ and $I$ represent capital stock and investment respectively. The initial csapital stock is the firm's fixed assets annual net average when it first appears in the sample, and $I_t$ is the deflated difference between this and previous period's fixed assets original value. Second, we estimate firms' TFP by the Levinsohn and Petrin method (Levinsohn & Petrin, 2003), where the deflated value added (in natural logarithm) is regressed against the number of workers and capital stock (both in natural logarithm) and intermediate inputs serves as a proxy variable to address the endogeneity problem in production function estimation.[27] The coefficients of labor and capital are estimated to be 0.1298 (standard error 0.0464) and 0.1888 (standard error 0.0624). Third, we predict TFP, divide it by $(w+m)$, and discretize into $K=4$ equal intervals, where state 1 is less than 25 percentile, state 2 is between 25 and 50 percentiles, state 3 is between 50 and 75 percentiles, and state 4 is above 75 percentile.

[27] The estimation utilizes the *levpet* package (Petrin, Poi, & Levinsohn, 2004) in Stata.

With the alternative measurement of states, we re-estimate the transition probability matrices, and Figures A3 and A4 in the appendix present the CDFs by innovation-export status and by states. Compared with Figures 5 and 6, we observe variations. Nevertheless, FOSD by states largely continues to hold, and FOSD of innovation- export against no-innovation-no-export continues to hold for states 1 and 4. We also re-estimate the dynamic decisions on whether to innovate and export, and Table 13 reports the results. While there exist variations in point estimates of coefficients, the pattern of large entry costs and small fixed costs continues to hold. Therefore, our findings appear largely robust to an alternative measurement of states.

Table 13. Estimation Results with Alternative Measurement of States

| | | Coef | S.E. | t |
|---|---|---|---|---|
| Constant | $\beta_0$ | -10.0376*** | 0.5483 | -18.3075 |
| Fixed cost of export | $\beta_1$ | -10.8404*** | 2.0886 | -5.1903 |
| Fixed cost of innovation | $\beta_2$ | -6.5202*** | 2.4921 | -2.6163 |
| Entry cost of export | $\beta_3$ | 7.4836*** | 0.6416 | 11.6642 |
| Entry cost of export, big firm | $\beta_4$ | -2.8712*** | 0.4886 | -5.8761 |
| Entry cost of innovation | $\beta_5$ | 6.7946*** | 0.5547 | 12.2494 |
| Entry cost of innovation, big firm | $\beta_6$ | -0.9519 | 0.5045 | -1.8868 |

Note: Standard errors are bootstrapped (50 replications); *** p<0.01; Fixed and entry costs are relative to fixed cost of production; N = 1089.

Source: The authors' estimation with data from NBS, China.

### *6.4 Processing trade*

A sizable proportion of exporting firms in the air-conditioner manufacturing industry is engaged in processing trade, where a firm imports all or part of the raw and auxiliary materials, parts, components, and packaging materials, and then re-exports the finished products after processing or assembly. Conceptually, it is likely that processing trade can affect a firm's product innovation. Via processing trade, a firm can learn how to manufacture a new product, which it then engineers for domestic market. To avoid such a possible confounding effect, and in light of the model fit in Table 8, we re-estimate the dynamic decisions on export and product innovation, excluding the processing trade firms.[28]

[28] Note the processing trade firms, by definition, import substantially. Hence by excluding them, we also control for the potential confounding effect of import.

Table 14. Dynamic Choice Model Estimation Results with Non-processing Trade Firms

| | | Coef. | S.E. | t |
|---|---|---|---|---|
| Constant | $\beta_0$ | -9.9765*** | 0.2703 | -36.9131 |
| Fixed cost of export | $\beta_1$ | 0.5822 | 0.6558 | 0.8877 |
| Fixed cost of innovation | $\beta_2$ | -1.0375 | 1.0766 | -0.9637 |
| Entry cost of export | $\beta_3$ | 8.3874*** | 0.6557 | 12.7918 |
| Entry cost of export, big firm | $\beta_4$ | -1.8235 | 1.1313 | -1.6118 |
| Entry cost of innovation | $\beta_5$ | 7.2490*** | 0.6445 | 11.2473 |
| Entry cost of innovation, big firm | $\beta_6$ | -2.1137 | 1.1046 | -1.9135 |

Note: Standard errors are bootstrapped (50 replications); *** p<0.01. Fixed and entry costs are relative to fixed cost of production. N = 864.

Source: The authors' estimation with data from NBS, China.

To do so, we first check the Customs data to see whether an exporting firm was engaged in processing trade. Firms are classified as processing trade firms if they participated in processing trade in at least one year during 2000, 2001, 2002, 2003, 2005 and 2006, which accounts for around 51% of the exporting firms. Then, we restrict the sample to non-processing trade firms, and re-estimate the dynamic choice model of export and product innovation. Table 14 reports the estimation results. Compared with Table 7, we can observe that, despite of difference in the magnitude of point estimates, the estimation results are consistent with those of Table 7. In particular, the pattern of large entry costs and small fixed costs continues to hold. Therefore, our finding is robust to the potential confounding effect of processing trade.

## 7. External Validity

Based on data from the Chinese air-conditioner manufacturing industry and exploiting China's WTO accession as a trade liberalization shock, we find a mechanism through which trade liberalization promotes product innovation, namely trade liberalization reduces trade costs, which in turn provides one static and two dynamic incentives for firms to innovate. Given the empirical focus on air-conditioner manufacturers in China from 2000 to 2007, the estimated quantitative effects should not be interpreted as universal responses to trade liberalization. Instead, the external validity of our analysis concerns whether the identified mechanism generalizes to other environments, namely across different industries/products, different countries/institutions, and different types of trade liberalization.

Industry-wise, our modelling features differentiated products, substantial opportunities to expand market size through exporting, and product innovation that increases firms' demand. In such industries, reductions in trade costs increase the return to innovation since marginal benefits of a successful product innovation are greater with a larger potential market. In contrast, in industries where products are relatively homogeneous, innovation is more process/marketing/organizational-oriented, or access to foreign markets provides only a limited increase in potential demand, the mechanism may be of less importance. Consequently, the mechanism identified here shall not be interpreted as implying that trade liberalization increases all forms of innovation to the same extent.

Institution-wise, China in the period from 2000 to 2007 features substantial structural transformation and rapid integration into the global markets. Its WTO accession was more a portfolio of trade policy shift that substantially reduced the barriers facing Chinese firms to access world markets, rather than mere tariff cuts. That is, it is more an extensive-margin trade liberalization, rather than an intensive-margin liberalization (namely trade liberalization for an already highly open economy). Consequently, the response to a comparable reduction in trade costs in an already highly open economy may be quantitatively different.

Regarding trade-liberalization types, the identified mechanism in this study is most directly related to reductions in export costs (and subsequently increase in probability of export). Consequently, the mechanism may not be applicable to all forms of trade liberalization, such as import liberalization, tariff reductions on intermediate inputs, and reductions in FDI restrictions, unless they also promote export participation.

In summary, the qualitative prediction that lower export costs can increase incentives of product innovation through one static and two dynamic channels shall apply in industries and economies where products are differentiated, trade liberalization is substantial and results in reduction in export costs, exporting expands the market available to innovative firms and innovation increases the profitability of serving that market. Nevertheless, the estimated magnitudes are likely to vary with market size, trade costs, innovation costs, productivity dynamics, and industry characteristics. That is, our main external-validity claim should be understood as mechanism-level rather than parameter-level.

## 8. Concluding Remarks

This study explores firms' optimal decisions to export and conduct product innovation both theoretically and empirically. Theoretically, we model firms' decisions on entry, exit, whether to innovate and export, the degree of product innovation, and price setting in both the domestic and export markets. The theoretical model features a mixed static and dynamic decision-making processes, and explicitly considers firms' entry and exit decisions.

Under the guidance of theoretical model, we then bring the model to a sample of air-conditioner manufacturing firms in China from 2000 to 2007, in order to estimate the effect of China's WTO accession in 2001. Such a trade liberalization shock allows us to estimate its effect on export and subsequently how the effect is transmitted to product innovation. We estimate that the change in the export environment associated with China's WTO accession is equivalent to a 13.5% reduction in iceberg trade cost, subsequently promoting firms' export. The export promotion then improves firms' product innovation through three mechanisms: one static channel and two dynamic channels.

Statically, trade liberalization promotes product innovation through a larger contemporaneous marginal benefit of innovation from a given reduction of iceberg trade cost for exporting firms, which incentivizes them to innovate. Dynamically, export and product innovation helps firms to predict a higher probability of transition to a better state in the future. In addition, as firms export and innovate, they can save on the entry costs in the future, which are substantial as per estimations.

Trade liberalization is generally considered to be beneficial in the sense of promoting international trade, particularly for developing economies. Moreover, our findings suggest that it can also promote innovation, both in the short and long terms. Such a benefit is to some extent an unintended consequence of trade liberalization.

In addition, the innovation effects of trade liberalization cannot be inferred from the contemporaneous expansion of market access alone. In our setting, a substantially large proportion of the innovation response operates through firms' forward-looking decisions: exporting and innovating alter the value of future productivity states and reduce future entry costs. Trade liberalization therefore changes not only the current profitability of innovation but also the dynamic incentives governing firms' subsequent investment decisions.

## References

Abbring, J. H. (2010). Identification of Dynamic Discrete Choice Models. *Annual Review of Economics, 2*(Volume 2, 2010), 367-394. doi:https://doi.org/10.1146/annurev.economics.102308.124349

Abbring, J. H., & Daljord, Ø. (2020). Identifying the discount factor in dynamic discrete choice models. *Quantitative Economics, 11*(2), 471-501. doi:https://doi.org/10.3982/QE1352

Aghion, P., Bergeaud, A., Lequien, M., & Melitz, M. J. (2018). The Heterogeneous Impact of Market Size on Innovation: Evidence from French Firm-Level Exports. *National Bureau of Economic Research Working Paper Series, No. 24600*. doi:10.3386/w24600

Aguirregabiria, V. (2010). Another Look at the Identification of Dynamic Discrete Decision Processes: An Application to Retirement Behavior. *Journal of Business & Economic Statistics, 28*(2), 201-218. doi:10.1198/jbes.2009.07072

Aguirregabiria, V., & Mira, P. (2010). Dynamic discrete choice structural models: A survey. *Journal of Econometrics, 156*(1), 38-67. doi:https://doi.org/10.1016/j.jeconom.2009.09.007

Alessandria, G., & Choi, H. (2014). Do falling iceberg costs explain recent U.S. export growth? *Journal of International Economics, 94*(2), 311-325. doi:https://doi.org/10.1016/j.jinteco.2014.08.009

Arcidiacono, P., & Miller, R. A. (2011). Conditional Choice Probability Estimation of Dynamic Discrete Choice Models With Unobserved Heterogeneity. *Econometrica, 79*(6), 1823-1867. doi:https://doi.org/10.3982/ECTA7743

Atkeson, A., & Burstein, A. (2010). Innovation, Firm Dynamics, and International Trade. *Journal of Political Economy, 118*(3), 433-484. doi:10.1086/653690

Aw, B. Y., Roberts, M. J., & Xu, D. Y. (2011). R&D Investment, Exporting, and Productivity Dynamics. *American Economic Review, 101*(4), 1312-1344. doi:doi: 10.1257/aer.101.4.1312

Ayllón, S., & Radicic, D. (2019). Product innovation, process innovation and export propensity: persistence, complementarities and feedback effects in Spanish firms. *Applied Economics, 51*(33), 3650-3664. doi:10.1080/00036846.2019.1584376

Bernard, A. B., & Jensen, B. J. (1999). Exceptional exporter performance: cause, effect, or both? *Journal of International Economics, 47*(1), 1-25. doi:https://doi.org/10.1016/S0022-1996(98)00027-0

Bernard, A. B., & Jensen, J. B. (2004). Why Some Firms Export. *The Review of Economics and Statistics, 86*(2), 561-569. Retrieved from http://www.jstor.org/stable/3211647

Bilir, L. K., & Morales, E. (2020). Innovation in the Global Firm. *Journal of Political Economy, 128*(4), 1566-1625. doi:10.1086/705418

Blalock, G., & Gertler, P. J. (2004). Learning from exporting revisited in a less developed setting. *Journal of Development Economics, 75*(2), 397-416. doi:https://doi.org/10.1016/j.jdeveco.2004.06.004

Bloom, N., Draca, M., & Van Reenen, J. (2015). Trade Induced Technical Change? The Impact of Chinese Imports on Innovation, IT and Productivity. *The Review of Economic Studies, 83*(1), 87-117. doi:10.1093/restud/rdv039

Blyde, J., Iberti, G., & Mussini, M. (2018). When does innovation matter for exporting? *Empirical Economics, 54*(4), 1653-1671. doi:10.1007/s00181-017-1274-x

Bratti, M., & Felice, G. (2012). Are Exporters More Likely to Introduce Product Innovations? *The World Economy, 35*(11), 1559-1598. doi:https://doi.org/10.1111/j.1467-9701.2012.01453.x

Bugni, F. A., & Ura, T. (2019). Inference in dynamic discrete choice problems under local misspecification. *Quantitative Economics, 10*(1), 67-103. doi:https://doi.org/10.3982/QE917

Bustos, P. (2011). Trade Liberalization, Exports, and Technology Upgrading: Evidence on the Impact of MERCOSUR on Argentinian Firms. *American Economic Review, 101*(1), 304-340. doi:10.1257/aer.101.1.304

Clerides, S. K., Lach, S., & Tybout, J. R. (1998). Is Learning by Exporting Important? Micro-Dynamic Evidence from Colombia, Mexico, and Morocco*. *The Quarterly Journal of Economics, 113*(3), 903-947. doi:10.1162/003355398555784

Coelli, F., Moxnes, A., & Ulltveit-Moe, K. H. (2022). Better, Faster, Stronger: Global Innovation and Trade Liberalization. *The Review of Economics and Statistics, 104*(2), 205-216. doi:10.1162/rest_a_00951

Dai, M., Huang, W., & Zhang, Y. (2020). Persistent effects of initial labor market conditions: The case of China's tariff liberalization after WTO accession. *Journal of Economic Behavior & Organization, 178*, 566-581. doi:https://doi.org/10.1016/j.jebo.2020.07.036

Dai, M., Huang, W., & Zhang, Y. (2021). How do households adjust to tariff liberalization? Evidence from China's WTO accession. *Journal of Development Economics, 150*, 102628. doi:https://doi.org/10.1016/j.jdeveco.2021.102628

Dai, M., Liu, H., & Lin, L. (2020). How innovation impacts firms' export survival: Does export mode matter? *The World Economy, 43*(1), 81-113. doi:https://doi.org/10.1111/twec.12847

Damijan, J. P., Kostevc, Č., & Polanec, S. (2010). From Innovation to Exporting or Vice Versa? *The World Economy, 33*(3), 374-398. doi:https://doi.org/10.1111/j.1467-9701.2010.01260.x

Das, S., Roberts, M. J., & Tybout, J. R. (2007). Market Entry Costs, Producer Heterogeneity, and Export Dynamics. *Econometrica, 75*(3), 837-873. doi:10.1111/j.1468-0262.2007.00769.x

De Fuentes, C., Niosi, J., & Peerally, J. A. (2021). Exploring innovation and export interplay in Canadian firms. *Economics of Innovation and New Technology, 30*(8), 786-806. doi:10.1080/10438599.2020.1786999

De Loecker, J. (2007). Do exports generate higher productivity? Evidence from Slovenia. *Journal of International Economics, 73*(1), 69-98. doi:https://doi.org/10.1016/j.jinteco.2007.03.003

De Loecker, J. (2013). Detecting Learning by Exporting. *American Economic Journal: Microeconomics, 5*(3), 1-21. doi:10.1257/mic.5.3.1

Dixit, A. K., & Pindyck, R. S. (1994). *Investment Under Uncertainty*. Princeton (NJ): Princeton University Press.

Ederington, J., & McCalman, P. (2008). Endogenous firm heterogeneity and the dynamics of trade liberalization. *Journal of International Economics, 74*(2), 422-440. doi:https://doi.org/10.1016/j.jinteco.2007.07.001

Egger, P., & Keuschnigg, C. (2015). Innovation, Trade, and Finance. *American Economic Journal: Microeconomics, 7*(2), 121-157. doi:doi: 10.1257/mic.20120032

Feng, L., Li, Z., & Swenson, D. L. (2017). Trade policy uncertainty and exports: Evidence from China's WTO accession. *Journal of International Economics, 106*, 20-36. doi:https://doi.org/10.1016/j.jinteco.2016.12.009

Hotz, V. J., & Miller, R. A. (1993). Conditional Choice Probabilities and the Estimation of Dynamic Models. *The Review of Economic Studies, 60*(3), 497-529. doi:10.2307/2298122

Imai, S., Jain, N., & Ching, A. (2009). Bayesian Estimation of Dynamic Discrete Choice Models. *Econometrica, 77*(6), 1865-1899. doi:https://doi.org/10.3982/ECTA5658

Impullitti, G., & Licandro, O. (2018). Trade, Firm Selection and Innovation: The Competition Channel. *The Economic Journal, 128*(608), 189-229. doi:https://doi.org/10.1111/ecoj.12466

Kalouptsidi, M., Scott, P. T., & Souza-Rodrigues, E. (2021). Identification of counterfactuals in dynamic discrete choice models. *Quantitative Economics, 12*(2), 351-403. doi:https://doi.org/10.3982/QE1253

Keane, M. P., & Sauer, R. M. (2009). Classification Error in Dynamic Discrete Choice Models: Implications for Female Labor Supply Behavior. *Econometrica, 77*(3), 975-991. doi:https://doi.org/10.3982/ECTA7642

Keane, M. P., & Wolpin, K. I. (2009). Empirical applications of discrete choice dynamic programming models. *Review of Economic Dynamics, 12*(1), 1-22. doi:https://doi.org/10.1016/j.red.2008.07.001

Levinsohn, J., & Petrin, A. (2003). Estimating Production Functions Using Inputs to Control for Unobservables. *The Review of Economic Studies, 70*(2), 317-341. doi:10.1111/1467-937x.00246

Lileeva, A., & Trefler, D. (2010). Improved Access to Foreign Markets Raises Plant-level Productivity for Some Plants. *The Quarterly Journal of Economics, 125*(3), 1051-1099. doi:10.1162/qjec.2010.125.3.1051

Lin, F., & Long, C. X. (2020). The impact of globalization on youth education: Empirical evidence from China's WTO accession. *Journal of Economic Behavior & Organization, 178*, 820-839. doi:https://doi.org/10.1016/j.jebo.2020.08.024

Lin, H.-l., & Lin, E. S. (2010). FDI, Trade, and Product Innovation: Theory and Evidence. *Southern Economic Journal, 77*(2), 434-464. doi:https://doi.org/10.4284/sej.2010.77.2.434

Liu, Q., & Ma, H. (2020). Trade policy uncertainty and innovation: Firm level evidence from China's WTO accession. *Journal of International Economics, 127*, 103387. doi:https://doi.org/10.1016/j.jinteco.2020.103387

Lu, Y., & Yu, L. (2015). Trade Liberalization and Markup Dispersion: Evidence from China's WTO Accession. *American Economic Journal: Applied Economics, 7*(4), 221-253. doi:10.1257/app.20140350

Melitz, M. J., & Redding, S. J. (2021). Trade and Innovation. *National Bureau of Economic Research Working Paper Series, No. 28945*. doi:10.3386/w28945

OECD, & Eurostat. (2005). *Oslo Manual: Guidelines for Collecting and Interpreting Innovation Data* (3rd Editio ed.).

Olabisi, M. (2017). The Impact of Exporting and Foreign Direct Investment on Product Innovation: Evidence from Chinese Manufacturers. *Contemporary Economic Policy, 35*(4), 735-750. doi:https://doi.org/10.1111/coep.12227

Petrin, A., Poi, B. P., & Levinsohn, J. (2004). Production Function Estimation in Stata using Inputs to Control for Unobservables. *The Stata Journal, 4*(2), 113-123. doi:10.1177/1536867x0400400202

Roberts, M. J., & Tybout, J. R. (1997). The Decision to Export in Colombia: An Empirical Model of Entry with Sunk Costs. *The American Economic Review, 87*(4), 545-564. Retrieved from http://www.jstor.org/stable/2951363

Rust, J. (1987). Optimal Replacement of GMC Bus Engines: An Empirical Model of Harold Zurcher. *Econometrica, 55*(5), 999-1033. doi:10.2307/1911259

Salomon, R. M., & Shaver, J. M. (2005). Learning by Exporting: New Insights from Examining Firm Innovation. *Journal of Economics & Management Strategy, 14*(2), 431-460. doi:https://doi.org/10.1111/j.1530-9134.2005.00047.x

Sun, S. (2023). Firm heterogeneity, worker training and labor productivity: the role of endogenous self-selection. *Journal of Productivity Analysis, 59*(2), 121-133. doi:10.1007/s11123-022-00652-1

Sun, S., & Anwar, S. (2024). Can We Reliably Identify the CES Preference Parameter from Firm Revenue and Cost Data? Evidence from Monte Carlo Experiments. *Journal of Quantitative Economics*. doi:10.1007/s40953-024-00397-8

Xiang, Y., Gubian, S., Suomela, B., & Hoeng, J. (2013). Generalized Simulated Annealing for Global Optimization: The GenSA Package. *The R Journal, 5*(1), 13-28. doi:10.32614/RJ-2013-002

# Appendix

## 1. Proofs

The following lemma is elementary, and may have been proved elsewhere. Nevertheless, we present it here for the sake of completeness. Its proof is due to ChatGPT.

**Lemma A1:** Let the following holds:

(1) $Z = \{z_1, z_2, \cdots\}$ is an ordered set with $z_1 < z_2 < \cdots$ that represent states;

(2) $\mu_1$ and $\mu_2$ are two transition probability matrices on $Z$ with $\mu_2$ first-order stochastically dominating $\mu_1$, namely for all $z_- \in Z$, $z \in Z$, $\mu_1(z|z_-) \in [0,1]$, $\mu_2(z|z_-) \in [0,1]$, $\sum_{z\in Z} \mu_1(z|z_-) = \sum_{z\in Z} \mu_2(z|z_-) = 1$ and $\sum_{z\geq z'} \mu_2(z|z_-) > \sum_{z\geq z'} \mu_1(z|z_-)$ for all $z' \in Z$;

(3) $g: Z \to \mathcal{R}$ is monotone increasing.

Then for all $z_- \in Z$, $\sum_{z\in Z} g(z)(\mu_2 - \mu_1)(z|z_-) \geq 0$.

*Proof*: Fix a $z_- \in Z$. For any probability distribution $\tilde{\mu}$ on $Z$, we have the following identity:

$$\sum_{i\geq 1} g(z_i)\tilde{\mu}(z_i) = g(z_1) + \sum_{i\geq 1}[g(z_{i+1}) - g(z_i)]\left(\sum_{j\geq i} \tilde{\mu}(z_j)\right).$$

Therefore, $\sum_{z\in Z} g(z)(\mu_2 - \mu_1)(z|z_-) = \sum_{i\geq 1}[g(z_{i+1}) - g(z_i)]\left(\sum_{j\geq i}\left(\mu_2(z_j|z_-) - \mu_1(z_j|z_-)\right)\right)$. Since $\mu_2$ first-order stochastically dominates $\mu_1$, we have $\sum_{j\geq i}\left(\mu_2(z_j|z_-) - \mu_1(z_j|z_-)\right) \geq 0$ for $i \geq 1$. Since $g(z)$ is monotone increasing, $g(z_{i+1}) - g(z_i) \geq 0$ for $i \geq 1$. Consequently, $\sum_{z\in Z} g(z)(\mu_2 - \mu_1)(z|z_-) \geq 0$.

□

**Proposition 1**: Static benefit of a reduction of iceberg trade cost
A decrease of the iceberg trade cost immediately incentivizes firms to innovate, should they export.

*Proof*: $\frac{\partial\tilde{\pi}(\chi_1=1)}{\partial\tau} = \frac{\partial\tilde{\pi}(\chi_1=0)}{\partial\tau}\left(1 + \lambda_2^{\frac{\rho}{1-\rho}}\zeta^{\frac{\rho}{1-\rho}}\right)^{\frac{\tilde{\rho}}{1-\tilde{\rho}}\frac{1-\rho}{\rho}} < \frac{\partial\tilde{\pi}(\chi_1=0)}{\partial\tau} < 0$. Hence, if firms export, the marginal benefit of a decrease in the iceberg trade cost is greater for innovating firms than non-innovating firms. This provides an immediate incentive for exporting firms to conduct product innovation.

□

**Proposition 2**: Dynamic benefits of a reduction of iceberg trade cost

(1) Saving of entry costs: If firms start to export and innovate after a reduction of iceberg trade cost, saving of the entry costs of export and innovation is a dynamic benefit.

(2) State transition: A reduction of iceberg trade cost brings in an additional dynamic benefit, should firms start to export and innovate and their belief of transition of productivity states is as follows:

(2.1) Export and innovation help them move to a higher productivity state, namely the transition probability matrix of $\left(\frac{\psi}{w+m}\right)$ when firms choose $(\chi_1, \chi_2) = (1,1)$ first-order stochastically dominates that of $(\chi_1, \chi_2) = (0,0)$;

(2.2) At a higher productivity state, they have a better chance of moving upward, namely the transition probability distribution of $\left(\frac{\psi}{w+m}\right)$ at a higher productivity state first-order stochastically dominates that of a lower productivity state.

*Proof*: For (1), by observing the value function, it is clear that $v_{(1,1)}$ is greater if firms export and innovate in the previous period.

For (2), first observe the per-period profit is a monotone increasing function of $\left(\frac{\psi}{w+m}\right)$. Second, the transition of productivity states $\left(\frac{\psi}{w+m}\right)$ is also monotone as per (2.2). Third, the feasible choice set does not depend on $\left(\frac{\psi}{w+m}\right)$. Fourth, $\delta\sigma < 1$ so that the Bellman operator is a contraction map. Consequently, the value function is monotone increasing function of $\left(\frac{\psi}{w+m}\right)$.

The continuation value $\mathbb{E}v_{(1,1)} = \sum v\left(\left(\frac{\psi}{w+m}\right)_+\right)\mu_{(1,1)}\left(\left(\frac{\psi}{w+m}\right)_+ \middle| \left(\frac{\psi}{w+m}\right)\right)$,

where the subscript $+$ represents next period and $\mu_{(1,1)}$ is the transition probability matrix of $\left(\frac{\psi}{w+m}\right)$ if firms choose $(\chi_1, \chi_2) = (1,1)$, and $\mathbb{E}v_{(0,0)} = \sum v\left(\left(\frac{\psi}{w+m}\right)_+\right)\mu_{(0,0)}\left(\left(\frac{\psi}{w+m}\right)_+ \middle| \left(\frac{\psi}{w+m}\right)\right)$ where $\mu_{(0,0)}$ is the transition probability matrix if $(\chi_1, \chi_2) = (0,0)$.

Therefore, given (2.1), by Lemma A1, we have $\mathbb{E}v_{(1,1)} - \mathbb{E}v_{(0,0)} \geq 0$. That is, firms' monotone belief results in an additional dynamic benefit of greater continuation value.

□

## 2. Tables and Figures

Table A1. The Multinomial Logit Modelling of Transition Probabilities

| | (1) First-stage bivariate probit | | (2) Second-stage MNL | | | (3) MNL | | |
|---|---|---|---|---|---|---|---|---|
| | $\chi_1$ | $\chi_2$ | 1 | 2 | 3 | 1 | 2 | 3 |
| $\chi_1$ | | | -1.285 | -0.913 | -1.724 | -1.330*** | -0.853** | -0.411 |
| | | | (1.926) | (1.848) | (1.283) | (0.509) | (0.404) | (0.325) |
| $\chi_2$ | | | -1.076 | 0.525 | 1.184 | -1.096*** | -0.859** | -0.513* |
| | | | (1.929) | (1.554) | (1.023) | (0.412) | (0.347) | (0.291) |
| dψ1 | -0.801*** | -1.282*** | 8.454 | 7.252 | 3.529 | 8.415*** | 6.589*** | 2.958*** |
| | (0.159) | (0.141) | (7.099) | (5.202) | (5.340) | (1.012) | (0.857) | (0.871) |
| dψ2 | -0.477*** | -0.985*** | 6.319 | 7.313*** | 5.025** | 6.285*** | 6.779*** | 4.519*** |
| | (0.145) | (0.134) | (4.780) | (2.275) | (2.105) | (0.896) | (0.686) | (0.573) |
| dψ3 | -0.122 | -0.747*** | 2.837 | 4.005*** | 4.426*** | 2.820*** | 3.582*** | 3.916*** |
| | (0.134) | (0.129) | (4.407) | (0.790) | (0.461) | (0.872) | (0.541) | (0.322) |
| eint | 0.351 | 0.810 | | | | | | |
| | (0.770) | (0.748) | | | | | | |
| npint | -0.242 | -0.994* | | | | | | |
| | (0.577) | (0.585) | | | | | | |
| sexp | -0.495 | 1.030*** | | | | | | |
| | (0.411) | (0.395) | | | | | | |
| sinn | 1.545*** | 0.538 | | | | | | |
| | (0.556) | (0.570) | | | | | | |
| sboth | 0.149 | 0.0808 | | | | | | |
| | (0.586) | (0.581) | | | | | | |
| cf1 | | | 0.0170 | 0.164 | 0.954 | | | |
| | | | (1.134) | (1.116) | (0.784) | | | |
| cf2 | | | -0.0768 | -0.935 | -1.230* | | | |
| | | | (1.174) | (0.969) | (0.636) | | | |
| Constant | -0.701*** | -0.109 | -3.767 | -3.869*** | -2.554*** | -3.728*** | -2.970*** | -1.788*** |
| | (0.149) | (0.142) | (4.458) | (1.387) | (0.816) | (0.734) | (0.486) | (0.287) |
| /athrho | 0.217*** | | | | | | | |
| | (0.0716) | | | | | | | |
| N | 798 | | 798 | | | 798 | | |

Note: In MNL estimations, state 4 is the base outcome; cf1 and cf2 are the generalized error terms, constructed from the first-stage bivariate probit estimation, that serve as control functions, and a chi-squared test of joint significance of cf1 and cf2 obtains a test statistic of 7.55 with a p-value of 0.2733; eint: average export intensity in the province-year, excluding a firm itself; npint: average share of new product value in sales in the province-year, excluding a firm itself; sexp: share of number of firms that export in the province-year, excluding a firm itself; sinn: share of number of firms that innovate in the province-year, excluding a firm itself; sboth: share of number of firms that both export and innovate in the province-year, excluding a firm itself; Bootstrap standard errors in parentheses (100 replications); *** $p<0.01$, ** $p<0.05$, * $p<0.1$.
Source: The author's estimation with data from NBS, China.

Figure A1. A Comparison with Paper Pulp Manufacturing Industry

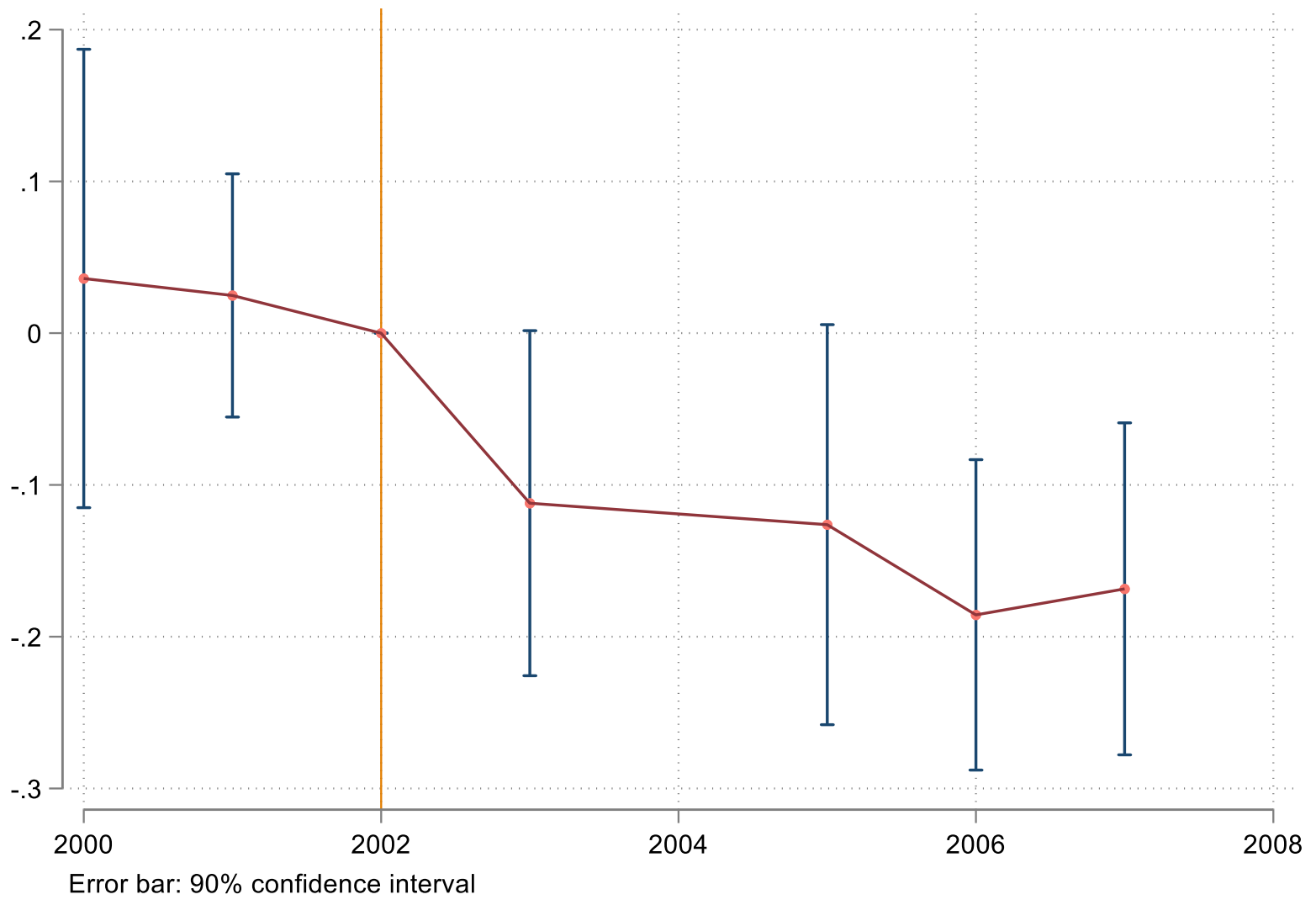

Figure A1. First Order Stochastic Dominance by Innovation and Export with Transition Probabilities Predicted from Multinomial Logit Model

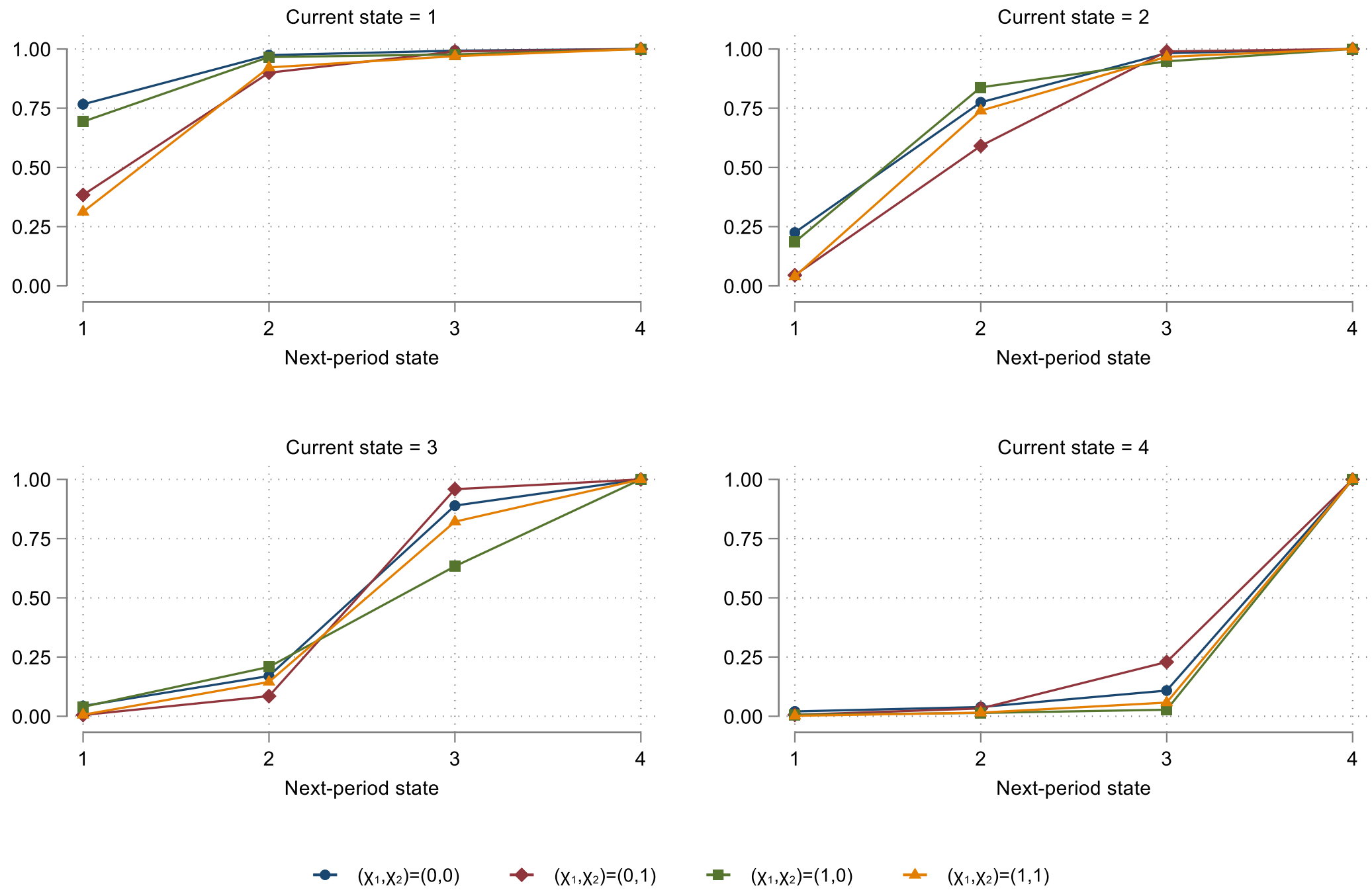

Figure A2. First Order Stochastic Dominance by Productivity States with Transition Probabilities Predicted from Multinomial Logit Model

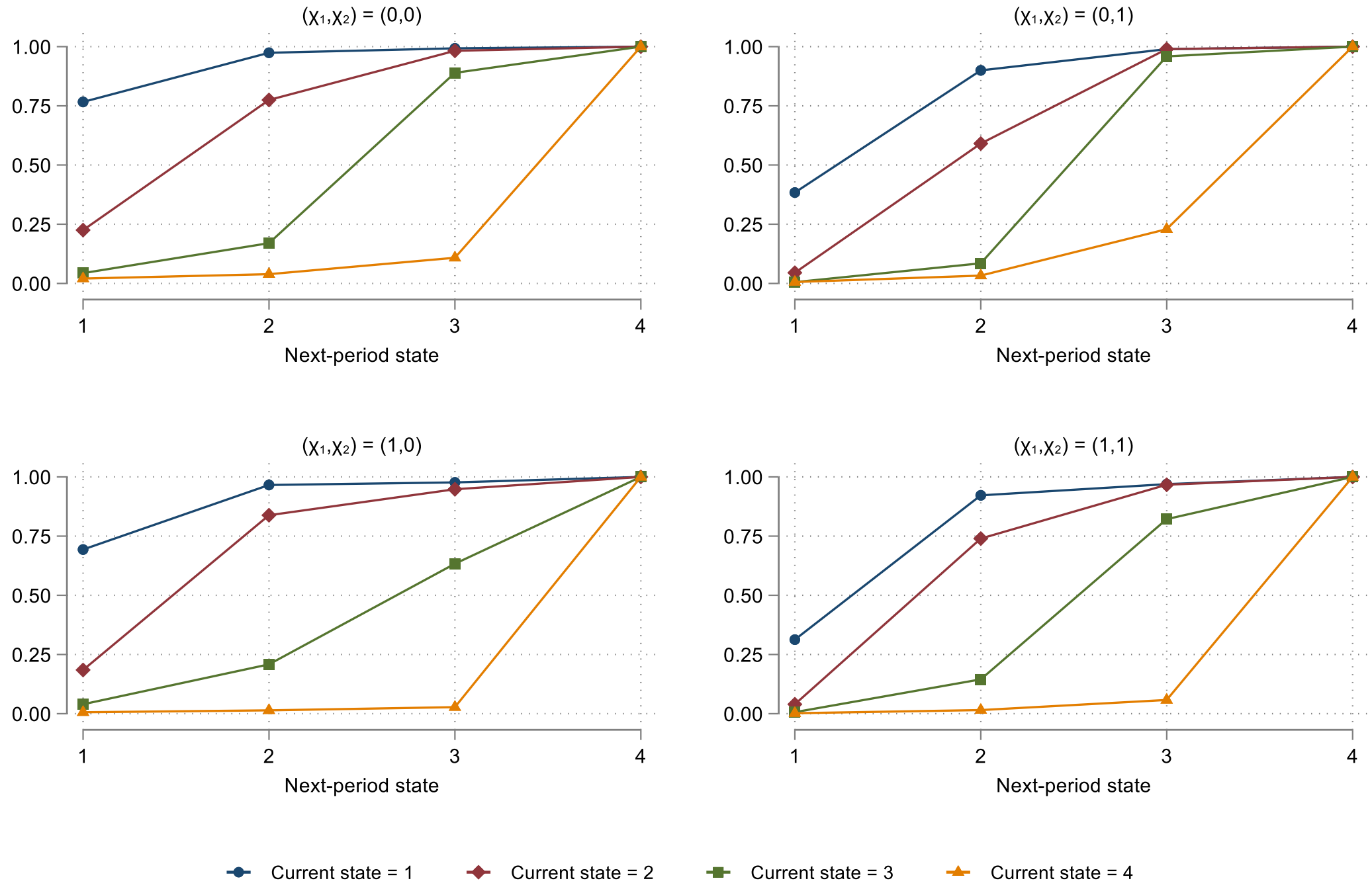

# Figure A3. First Order Stochastic Dominance by Innovation and Export with Alternative Measurement of States

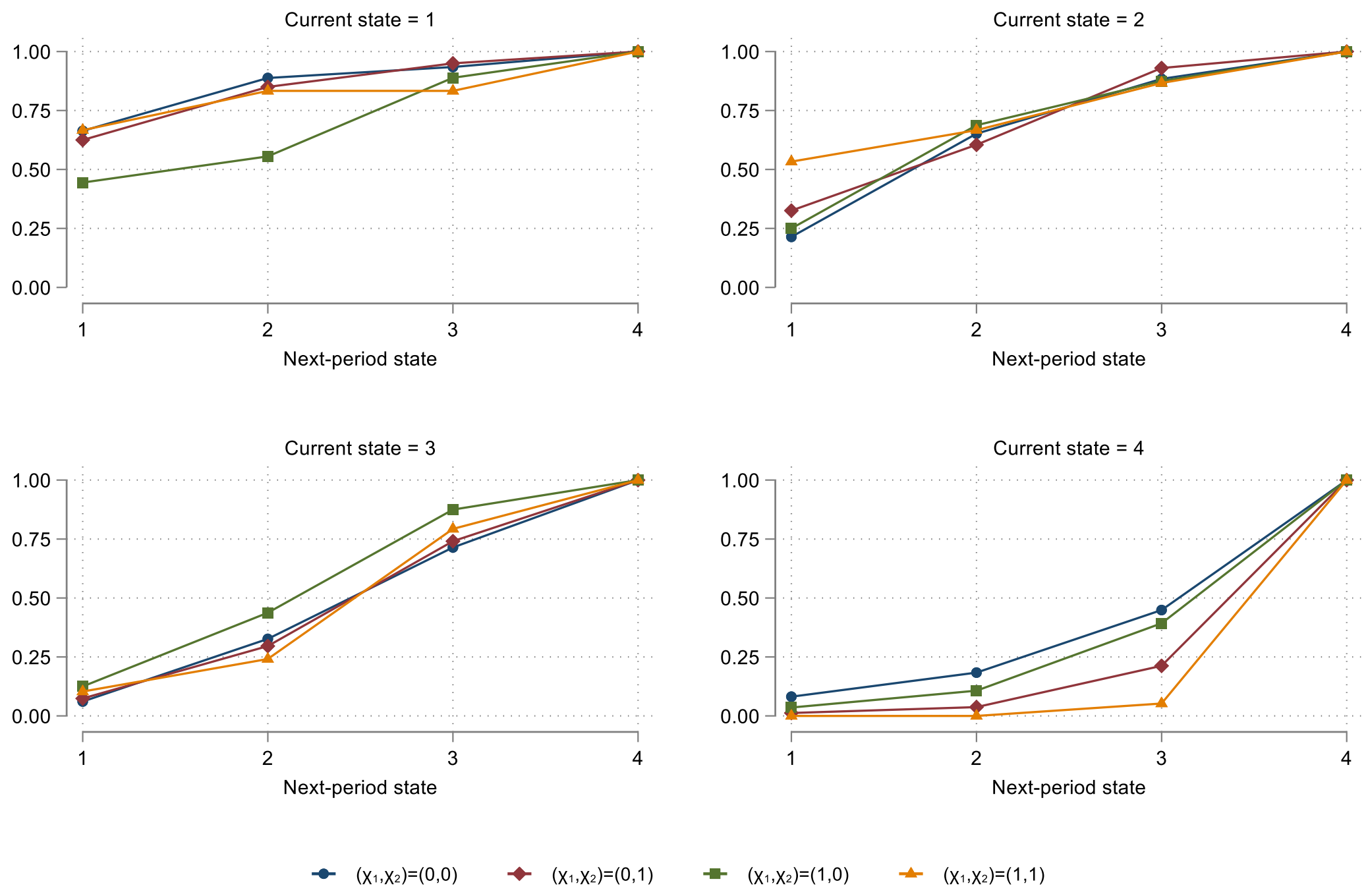

# Figure A4. First Order Stochastic Dominance by Productivity States with with Alternative Measurement of States

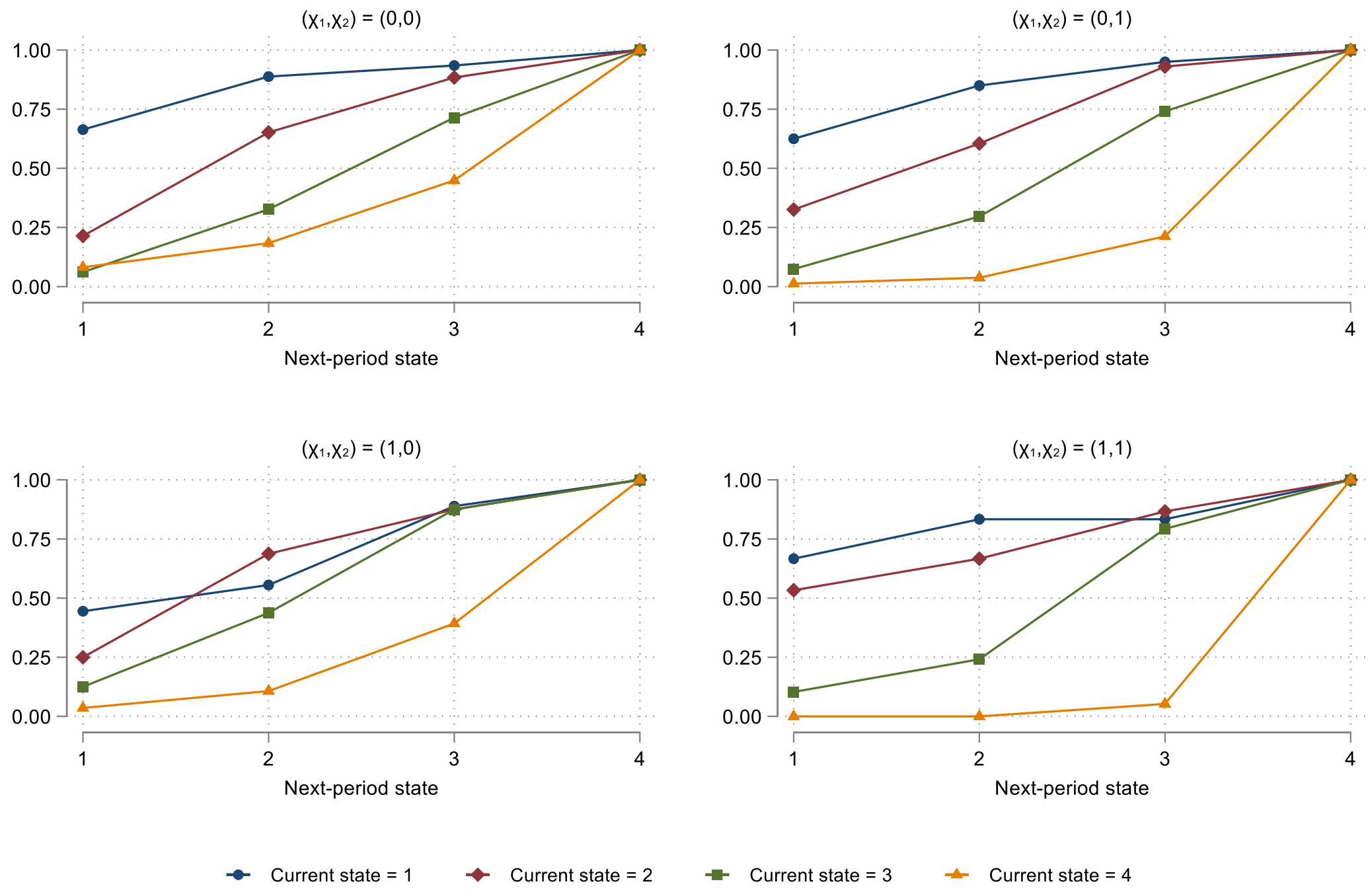